\documentclass[fleqn,usenatbib]{mnras}

\usepackage{newtxtext,newtxmath}

\usepackage[T1]{fontenc}
\usepackage{ae,aecompl}
\usepackage{subfig}
\usepackage{mwe}
\usepackage{xcolor}

\usepackage{graphicx}	
\usepackage{amsmath}	
\usepackage{amssymb}	
\usepackage{bm}

\newcommand{\dt}{\bm{\Delta t}}	

\newcommand{\sdt}{\Delta t}

\newcommand{\mt}{\bm{t_{\tilde{m}_{k}}}}
\newcommand{\mtW}{\bm{t}_{W,\tilde{\textbf{m}}_{\textbf{k}}}}
\newcommand{\mtM}{\bm{t}_{M,\tilde{\textbf{m}}_{\textbf{k}}}}
\newcommand{\mtS}{\bm{t}_{S,\tilde{\textbf{m}}_{\textbf{k}}}}
\newcommand{\Ddt}{{D_{\Delta t}}}
\newcommand{\micro}{\tilde{\textbf{m}}_{\textbf{k}}}
\newcommand{\macro}{\tilde{\textbf{M}}}
\newcommand{\macrop}{\bm{\xi_{\tilde{M}}}}
\newcommand{\sref}[1]{Section~\ref{#1}}
\newcommand{\fref}[1]{Figure~\ref{#1}}
\newcommand{\tref}[1]{Table~\ref{#1}}
\newcommand{\eref}[1]{Equation~(\ref{#1})}
\newcommand{\ebref}[1]{Equation~\ref{#1}}
\newcommand\pg{PG\,1115$+$080}
\newcommand\rxj{RXJ\,1131$-$1231}

\def\zd{z_{\rm d}}
\def\zs{z_{\rm s}}

\title[Constraining the microlensing effect in $H_{0}$ measurements]{Constraining the microlensing effect on time delays with new time-delay prediction model in $H_{0}$ measurements}

\author[G.~C.-F.~Chen et al.]{
Geoff~C.-F.~Chen,$^{1}$\thanks{E-mail: chfchen@ucdavis.edu}
James~H.~H.~Chan,$^{2}$
Vivien~Bonvin,$^{2}$
\newauthor{
Christopher~D.~Fassnacht,$^{1}$
Karina Rojas,$^{3,4}$
Martin~Millon,$^{2}$
Fred~Courbin,$^{2}$
}
\newauthor{
Sherry~H.~Suyu,$^{5,6,7}$
Kenneth~C.~Wong,$^{8}$
Dominique Sluse,$^{9}$
Tommaso Treu,$^{10}$
}
\newauthor{
Anowar J. Shajib,$^{10}$
Jen-Wei Hsueh,$^{1}$
David~J.~Lagattuta,$^{11}$
}
\newauthor{
L$\acute{\textrm{e}}$on~V.~E.~Koopmans,$^{12}$
Simona Vegetti,$^{5}$
and John~P.~McKean$^{12,13}$
}
\\
$^{1}$Department of Physics, University of California, Davis, CA 95616, USA\\
$^{2}$Institute of Physics, Laboratoire d'Astrophysique, Ecole Polytechnique F$\acute{\text{e}}$d$\acute{\text{e}}$rale de Lausanne (EPFL), Observatoire de Sauverny,\\~~CH-1290 Versoix, Switzerland\\
$^{3}$Instituto de F\'isica y Astronom\'ia, Universidad de Valpara\'iso, Avda. Gran Breta\~na 1111, Playa Ancha, Valpara\'iso 2360102, Chile \\
$^{4}$LSSTC Data Science Fellow \\
$^{5}$Max Planck Institute for Astrophysics, Karl-Schwarzschild-Strasse 1, D-85740 Garching, Germany\\
$^{6}$Institute of Astronomy and Astrophysics, Academia Sinica (ASIAA), P.O. Box 23-141, Taipei 10617, Taiwan\\
$^{7}$Physik-Department, Technische Universit\"{a}t M\"{u}nchen, James-Franck-Stra{\ss}e 1, 85748 Garching, Germany\\
$^{8}$National Astronomical Observatory of Japan, 2-21-1 Osawa, Mitaka, Tokyo 181-8588, Japan\\
$^{9}$STAR Institute, Quartier Agora~--~All$\acute{\text{e}}$e du six A$\hat{\text{o}}$ut, 19c, B-4000 Li$\grave{\text{e}}$ge, Belgium\\
$^{10}$Department of Physics and Astronomy, University of California, Los Angeles, CA 90095-1547, USA\\
$^{11}$CRAL, Observatoire de Lyon, Universit Lyon 1, 9 Avenue Ch. Andr, F-69561 Saint Genis Laval Cedex, France\\
$^{12}$Kapteyn Astronomical Institute, University of Groningen, PO Box 800, NL-9700 AV Groningen, the Netherlands\\
$^{13}$Netherlands Institute for Radio Astronomy (ASTRON), PO Box 2, NL-7990 AA Dwingeloo, the Netherlands\\
}

\date{Accepted XXX. Received YYY; in original form ZZZ}

\pubyear{2018}
\begin{document}
\label{firstpage}
\pagerange{\pageref{firstpage}--\pageref{lastpage}}
\maketitle

\begin{abstract}
Time-delay strong lensing provides a unique way to directly measure the Hubble constant ($H_{0}$). The precision of the $H_{0}$ measurement depends on the uncertainties in the time-delay measurements, the mass distribution of the main deflector(s), and the mass distribution along the line of sight. 
\citet{TieKochanek18} have proposed a new microlensing effect on time delays based on differential magnification of the coherent accretion disc variability of the lensed quasar. 
If real, this effect could significantly broaden the uncertainty on the time delay measurements by up to 30\% for lens systems such as \pg, which have relatively short time delays and monitoring over several different epochs.
In this paper we develop a new technique that uses the cosmological time-delay ratios and simulated microlensing maps within a Bayesian framework in order to limit the allowed combinations of microlensing delays and thus to lessen the uncertainties due to the proposed effect.
We show that, under the assumption of \citet{TieKochanek18}, the uncertainty on the time-delay distance ($\Ddt$, which is proportional to 1/$H_{0}$) of short time-delay ($\sim 18$ days) lens, \pg, increases from $\sim 7$\% to $\sim 10$\% by simultaneously fitting the three time-delay measurements from the three different datasets across twenty years, while in the case of long time-delay ($\sim 90$ days) lens, the microlensing effect on time delays is negligible as the uncertainty on $\Ddt$ of \rxj~only increases from $\sim 2.5$\% to $\sim 2.6$\%. 
\end{abstract}

\begin{keywords}
gravitational lensing: micro -- cosmology: distance scale -- methods: data analysis
\end{keywords}



\section{Introduction}
The standard flat $\Lambda$CDM model has become a concordance cosmological model, which assumes spatial flatness, a matter content dominated by cold dark matter, and an accelerated expansion caused by dark energy \citep{Planck16a}.
Intriguingly, even though the standard flat $\Lambda$CDM model provides an excellent fit to various large-scale observables, including the cosmic microwave background (CMB) and Baryon Acoustic Oscillations \citep[BAO;][]{KomatsuEtal11,HinshawEtal13}, 
the current $\sim3\sigma$ tension between direct measurements of $H_{0}$ and that inferred from Planck data based on the flat $\Lambda$CDM model
may indicate new physics beyond the standard cosmological model \citep{RiessEtal16,Freedman17}. 
Therefore, to clarify whether this tension is due to systematics, multiple independent methods with 
precise (1\% or better) and accurate $H_{0}$ measurements 
are crucial for testing the possible hidden biases in any individual method \citep[e.g.,][]{Suyu12,WeinbergEtal13}. 

Time-delay strong lensing (TDSL), which uses gravitational lens systems in which a foreground galaxy produces multiple images of a variable background object such as a quasar, is a powerful technique for measuring $H_{0}$. Compared with Type-Ia supernovae, which need to be calibrated by either by distance ladder techniques \citep{RiessEtal98} or by an inverse distance ladder from BAO and CMB to yield $H_{0}$ \citep{AuborugEtal15}, TDSL is not only a completely independent method but also a one-step way to probe $H_{0}$.  The measurements are obtained by constraining the combined cosmological distances (or so-called time-delay distance, $\Ddt$), which are mostly sensitive to $H_{0}$ \citep[see the review by][]{TreuMarshall17}. 
While this method was proposed by \citet{Refsdal64} over fifty years ago, it is only in the last fifteen years 
that robust measurements of high enough precision have been achieved, recently yielding a 3.8\% accurate measurement of $H_{0}$ based on the time-delay measurements in three lenses \citep{BonvinEtal17}. 

The methodology of TDSL relies on three inputs for each lens: 
(1) multi-year lens monitoring programs to measure high-precision time delays \citep[e.g.,][]{FassnachtEtal02,TewesEtal13b,BonvinEtal17,RathnaEtal13,EulaersEtal13,CourbinEtal18}, 
(2) high resolution imaging and stellar kinematics to determine the mass distribution in the lensing galaxy \citep[e.g.,][]{TreuKoopmans02,KoopmansEtal03,SuyuEtal10,WongEtal17}, and 
(3) spectroscopy and multiband imaging to provide an inference of the mass distribution along the line of sight of the lens system \citep{SuyuEtal10,FassnachtEtal11,RusuEtal17,TihhonovaEtal17}. 
The error budget of each component, assuming they are independent, can be approximately translated to the $H_{0}$ error budget by $\sigma_{H_{0}}^2/H_{0}^2\propto(\sigma^{2}_{\delta t}/\delta t^2+\sigma_{\kappa}^{2}+\sigma_{\textrm{los}}^{2})/N$, where $\sigma_{\delta t}$, $\sigma_{\kappa}$, and $\sigma_{\textrm{los}}$ are the uncertainties on the time-delay measurements, the mass distribution of the main deflector(s), and the mass along the line of sight, respectively, and $N$ is the number of lenses. 
Since each lens is independent of another lens, one can keep pushing down the precision of the $H_{0}$ measurements by combining more and more lenses until one hits the systematic error floor in any individual component.
Current large sky surveys combining with numerical lens-finding techniques \citep[e.g.][]{JosephEtal14,AvestruzEtal17,Agnello17,PetrilloEtal17,OstrovskiEtal17,LanusseEtal18}, have already shown promising results and discovered many new lenses \citep[e.g.][]{LinEtal17,AgnelloEtal17, SchechterEtal17,OstrovskiEtal18,WilliamEtal18}. Furthermore, \citet{OguriMarshall10} forecast that we will discover thousands of lensed quasars with the Large Synoptic Survey Telescope.
Hence, a 1\% $H_{0}$ measurement is a realistic expectation in the near future \citep[e.g.,][Jee et al. 2018 submitted]{JeeEtal15,JeeEtal16,deGrijsEtal17,SuyuEtal18,ShajibEtal18} if we can control the systematic effects in each error budget to a sub-percent level. 

There are in general two ways to reveal systematic uncertainties. The first is performing a mock dataset challenge: mock datasets that mimic real data are created 
and then modelers analyze the  datasets and compare their results with truth to reveal any systematic effect in their modeling algorithms. 
For example, the  public time-delay challenge \citep[TDC,][]{DoblerEtal13,LiaoEtal15} aimed to examine the accuracy of different time-delay curve-fitting algorithms. The main purpose of the TDC was to understand how well we can control systematics on $\sigma_{\delta t}$. 
The conclusion was that if the measured time delay is the standard cosmological delay (see the definition in \ebref{eq:TDsum}) used in all lens models, it is feasible to measure accurate and precise time delays within 1\% \citep{TewesEtal13a,LiaoEtal15,BonvinEtal17}. 
Similarly, the on-going public time delay lens modeling challenge \citep[TDLMC,][]{DingEtal18} aims to test the accuracy of lens imaging modeling algorithms based on different source reconstruction techniques \citep[e.g.,][]{WarrenDye03,Koopmans05,VegettiKoopmans09,OguriAlgorithm10,NightingaleDye15,BirrerEtal15}. 
Additionally, the TDLMC may shed light on how critical the mass-sheet transformation (MST), a special case of source-position transformation, is \citep{FalcoEtal85,SchneiderSluse13,SchneiderSluse14,XuEtal15,BirrerEtal17_LOS}.
All in all, the goal of TDLMC is to 
understand how well we can control the systematic effects on $\sigma_{\kappa}$. 
In addition, \citet{BirrerEtal15} used mock data to study whether we can use lens imaging to detect small perturbations on $\sigma_{\kappa}$, while \citet{GChenEtal16} used mock data to study the impact of the unstable PSF on $\sigma_{\kappa}$ when using adaptive-optics imaging to study $H_{0}$.
However, it is difficult for mock dataset challenges to reveal the systematics caused by unknown physical phenomena because the mock data only include known processes. 

The second method to assess systematic effects is to study physical processes which have not been previously considered. 
For example, \citet[][hereafter TK18]{TieKochanek18} have questioned the use of measured time delays in cosmography, by showing that, under the assumption of the ``lamp-post'' model for accretion discs
and differential magnification of the disc stars in the lensing galaxy \citep[i.e. microlensing][]{Wambsganss06}, the measured time delays may introduce a bias in the inferred value of $H_{0}$.  
Under this assumed disc model, regions of the accretion disc that are separated by distances on order light days vary in a coherent manner in response to activity in the centre of the disc.
Differential magnification of portions of such a disc can introduce a phase delay due to the distance from the centre of the disk, and can shift the time-delay light curves by up to days depending on the accretion disk configuration and the microlensing pattern (see more description in \sref{sec:TDmap}).
Since each lensed image has a different microlensing pattern, the sum of this proposed microlensing time-delay effect for any pair of lensed images can be non-zero.
Therefore, the time delays we measure are not only the cosmological time delays but the combination of cosmological time delays and microlensing time delays,
\begin{equation}
\label{eq:first_TDsum}
\Delta t_{\textrm{measured}}=\Delta t_{\textrm{cosmological}}+\Delta t_{\textrm{microlensing}}.
\end{equation}
This effect, under certain assumptions, can significantly broaden the uncertainty on time-delay measurements, since it is embedded in the time-delay light curves (see Fig. 10 in TK18). 
Although long-term monitoring can partially average out and mitigate this microlensing effect on time delays, the non-zero mean cannot be removed (see Table. 2 in TK18). Thus, TK18 have claimed that the current uncertainty of $H_{0}$ measurements from TDSL could potentially be underestimated and biased. 
\begin{figure*}
\includegraphics[width=0.96\textwidth]{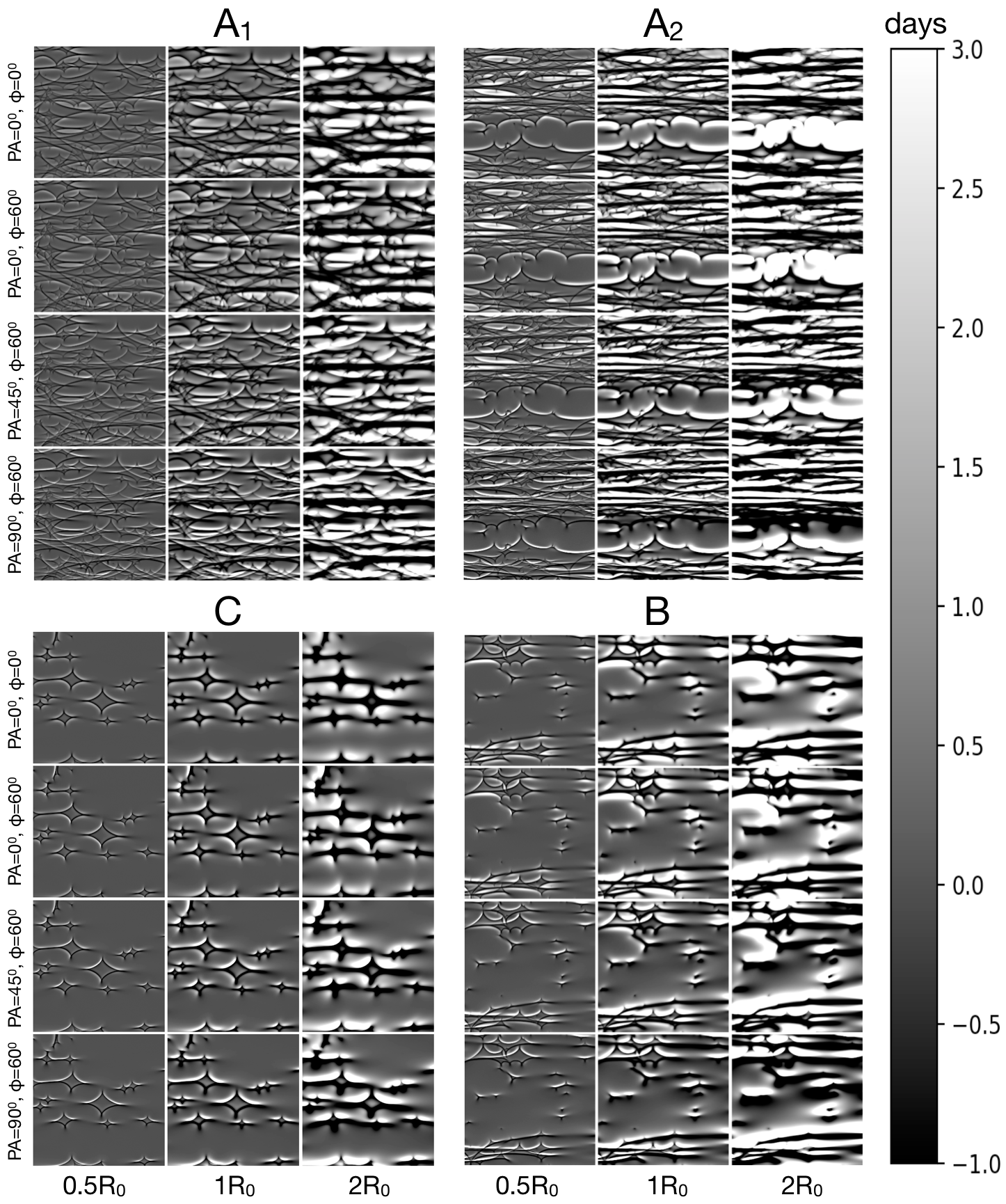}
\caption{The microlensing time-delay maps for image $A_{1}$ (top left panel), $A_{2}$ (top right panel), $B$ (bottom right panel), and $C$ (bottom left panel) of \pg. 
For each panel, the top row is for a face-on disc and the lower three rows are for a disc inclined by $\phi = 60^{\circ}$ with position angles of PA = $0^{\circ}$, $45^{\circ}$ and $90^{\circ}$, respectively. Each column refers to a different source size: (left) $0.5R_{0}$, (middle) $1R_{0}$ and (right) $2R_{0}$, where $R_{0}= 1.629\times10^{15}$ cm in the WFI $R_{c}$ filter (6517.25\AA) for an Eddington ratio of $L/L_{E}$ = 0.1 and a radiative efficiency of $\eta =0.1$, given an estimated black hole mass of $1.2 \times 10^{9}M_{\sun}$ from \citet{PengEtal06}.
All images are on the same scale with the minimum set at $-1$ day and maximum at $+3$ days, although certain pixels have delays that fall outside of this range. 
Black is used for negative delays and white for positive delays. Each map has the size of $20R_{\textrm{Ein}}$ with a 8192-pixel resolution, where $R_{\textrm{Ein}}=3.618 \times 10^{16}\textrm{cm}$ is the Einstein radius of a mean mass of the microlenses, $\langle M\rangle=0.3M_{\sun}$.}
\label{figure:TDmap}
\end{figure*}

Although the effect suggested by TK18 depends on an AGN variability model that is not yet well constrained observationally, in this paper we conservatively assume that this effect exists and 
develop a new technique to mitigate its consequences. 
We use a time-delay prediction model that incorporates the information from the cosmological time-delay ratio, which was first proposed by \citet{KeetonEtal09}\footnote{The extra time delays caused by substructures are negligible ($<0.5$ days) even when the mass of the substructures are larger than $10^{9}M_{\sun}$ \citep[see Fig. 3 in][]{KeetonEtal09,MaoSchneider98}.}, as well as the information from the microlensing time-delay maps.
In \sref{sec:TDmap}, we show the microlensing time-delay maps resulting from different source configurations. 
In \sref{sec:bayes}, we demonstrate how to properly infer $\Ddt$ by including the new microlensing effects using Bayesian inference\footnote{We use \texttt{emcee} \citep{Foreman-MackeyEtal13}, an MIT licensed pure-Python implementation of \citet{GoodmanWeare10} Affine Invariant Markov chain Monte Carlo (MCMC) Ensemble sampler, to perform the MCMC analysis. All the chains have converged based on the criteria in \citet{Foreman-MackeyEtal13}.} 
We show the time-delay modeling results of \pg~and \rxj~in \sref{sec:TDmodeling},
and summarize in \sref{sec:conclusion}. 
Note that, throughout the paper\footnote{We use \texttt{ChainConsumer}, a package developed by \citet{Hinton16}, to create color-blind accessible figures.}, we use the phrase ``microlensing time-delay effect'' to refer the microlensing effect on time delays proposed by TK18, and use ``microlensing magnification effect'' to refer the ``standard'' microlensing magnification of the image fluxes. 

\begin{table}
\caption{The $\kappa$, $\gamma$, and $\kappa_{\star}/\kappa$ at each lensed image position in \pg\ from the best fit of the macro model (Chen et al. 2018b in prep).  Values for \rxj~are taken from TK18.
}
\centering
\label{table:macroinput}
 \begin{tabular}{||c c c c c||} 
 \hline\hline
 Lens & Image & $\kappa$ & $\gamma$ & $\kappa_{\star}/\kappa$ \\ [0ex]
 \hline
 \pg & $A_{1}$ & 0.424 & 0.491 &0.259\\
     & $A_{2}$ & 0.451 & 0.626 &0.263\\
     & B & 0.502 & 0.811 &0.331\\
     & C & 0.356 & 0.315 & 0.203\\[0ex]  
 \hline
\end{tabular}
\end{table}

\section{microlensing time-delay maps}
\label{sec:TDmap}
In order to assess the magnitude of the microlensing time delay effect and to test our procedure, we need to create realizations of microlensing maps that are due to the stars in the lensing galaxies.  Rather than showing magnification, as is typical when showing microlensing realizations, these maps show the additional time delays introduced by the microlensing, under the assumption that the lamp-post model (see TK18) is correct.  
The details of creating the microlensing time-delay map for \pg~can be found in Bonvin et al. 2018 (hereafter B18). We summarize the key information in the following. We follow TK18 to produce microlensing time-delay maps at each lensed image position in a lens system given the total convergence ($\kappa$), the ratio of stellar convergence to total convergence ($\kappa_{\star}/\kappa$), and the shear ($\gamma$) from the best fit of the macro model\footnote{The reduced $\chi^{2}$ of the entire lens imaging modeling is $\approx 1$. 
The details of the \pg~lens imaging modeling will be presented in Chen et al. 2018b in prep.} (see \tref{table:macroinput} for \pg~and TK18 for \rxj). 
We assume a mean mass of the microlenses of $\langle M\rangle=0.3M_{\sun}$ following the Salpeter mass function with a ratio of the upper to lower masses of r = 100 \citep{KochanekDalal04}, although the choice of the mass function has little influence on our results (B18). 
We consider a standard thin disc model \citep{ShakuraSunyaev73} for the accretion disc given an estimated black hole mass of, e.g., $1.2\times10^{9}M_{\sun}$ for \pg\ \citep{PengEtal06}.
According to TK18, the microlensing screen due to the lensing galaxy may cause differential magnification of the accretion disk region of the background quasar. 
This can change the relative contributions of different parts of the accretion disk to the integrated flux of the image, and consequently change the average radius at which the variability takes place.
There are two main sources of the delay: (1) if the temperature profile (and hence brightness profile) of the disk responds to variations in the centre which then propagate outward through the accretion disk, the differentially magnified UV/optical emission from the disc can shift the light curve to a later time and also change its shape, (2) if the disk is tilted with respect to the line of sight, then there are extra light travel times from different parts of the disk.
We show the time-delay maps 
in \fref{figure:TDmap}, and list the combinations of different accretion disc sizes ($0.5R_{0}$, $1R_{0}$, and $2R_{0}$), different $\phi$ ($0^{\circ}$ and $60^{\circ}$), and different PA ($0^{\circ}$, $45^{\circ}$, and $90^{\circ}$), where $\phi$\footnote{Note that TK18 and B18 use $i$ to represent the inclination angle.} and PA represent the inclination and position angle of the disk with respect to the source plane, taken as perpendicular to the observer's line of sight ($\phi = 0$ corresponding to the face-on disc; see TK18 for a detailed explanation of the coordinates system).
The probability distribution of the time-delay maps with different combinations can be found in Fig. 5 of B18.

\section{Bayesian inference}
\label{sec:bayes}
In this section, we describe how we include and constrain the microlensing effects on time delays and properly infer $\Ddt$ under a Bayesian framework.
We denote $\dt$~as the measured time delays in \eref{eq:first_TDsum}, $\bm{d}$~as the lens imaging data,
$\micro$~as the microlensing model with a particular accretion disc property $\bm{k}$ (i.e., a particular combination of disc size, $\phi$, and PA),
$\mt$~as the parameters of the extra time delays at each lensed image caused by the microlensing model, $\macro$~as the macro model which is constrained by the lens imaging, $\macrop$ as the parameters of the macro model, and again $\Ddt$~as the time-delay distance.

The posterior of $\Ddt$, $\mt$, and $\macrop$~is 
\begin{gather}
\label{eq:poste}
\begin{align}
\textrm{P}&(\Ddt,\mt,\macrop|\dt,\bm{d},\micro,\macro)\nonumber\\
&\propto \textrm{P}(\dt,\bm{d}|\Ddt,\mt,\macrop,\micro,\macro)\nonumber\\
&~~~\cdot \textrm{P}(\Ddt)\textrm{P}(\mt|\micro,\macro)\textrm{P}(\macrop|\macro),
\end{align}
\end{gather}
where $\textrm{P}(\dt,\bm{d}|\Ddt,\mt,\macrop,\micro,\macro)$ 
is the joint likelihood of the lens and
\begin{equation}
\label{eq:prior}
\textrm{P}(\mt|\micro,\macro)=\prod_{i}^{N_{\textrm{im}}}\textrm{P}(t_{i,\micro}|\micro,\macro),
\end{equation} 
is the prior from the time delay maps in the microlensing model with a particular accretion disc property given the mass distribution from the macro model, $t_{i,\micro}$ are the extra time delays caused by the microlensing effect at the location of each lensed image $i$, and $N_{\textrm{im}}$ is the number of lensed images. Since the data are independent, we can decouple the joint likelihood as 
\begin{gather}
\label{eq:jointlikeli}
\begin{align}
\textrm{P}&(\dt,\bm{d}|\Ddt,\mt,\macrop,\micro,\macro)\nonumber\\
&=\textrm{P}(\dt|\Ddt,\mt,\macrop,\micro,\macro)\textrm{P}(\bm{d}|\macrop,\macro).
\end{align}
\end{gather}
We can substitute \eref{eq:jointlikeli} into \eref{eq:poste} and get
\begin{gather}
\label{eq:poste_new}
\begin{align}
\textrm{P}&(\Ddt,\mt,\macrop|\dt,\bm{d},\micro,\macro)\nonumber\\
&\propto \textrm{P}(\dt|\Ddt,\mt,\macrop,\micro,\macro)\nonumber\\
&~~~\cdot \textrm{P}(\Ddt)\textrm{P}(\mt|\micro,\macro)\textrm{P}(\bm{d}|\macrop,\macro)\textrm{P}(\macrop|\macro)\nonumber\\
&\approx \textrm{P}(\dt|\Ddt,\mt,\macrop,\micro,\macro)\nonumber\\
&~~~\cdot \textrm{P}(\Ddt)\textrm{P}(\mt|\micro,\macro)\textrm{P}(\macrop|\bm{d},\macro),
\end{align}
\end{gather}
where the likelihood, assuming a Gaussian distribution, can be expressed as
\begin{gather}
\label{eq:likelihood}
\begin{align}
\textrm{P}&(\dt|\Ddt,\mt,\macrop,\micro,\macro)\nonumber\\
&=\prod_{i,i<j}^{N_{\textrm{im}}}\frac{1}{\sqrt[]{2\pi}\sigma_{\sdt_{ij}}}\textrm{exp}\left[-\frac{(\sdt_{ij}-\sdt_{ij,\micro}^{\textrm{P}})^{2}}{2\sigma_{\sdt_{ij}}^{2}}\right],
\end{align}
\end{gather}
$j$ represents the reference lensed image in the time-delay modeling\footnote{Using the full covariance matrix of time-delay measurements is still under development, and beyond the scope of this paper.}, $\sdt_{ij}$ represents the measured time delays between lensed images $i$ and $j$, $\sdt_{ij,\micro}^{\textrm{P}}$ represents the predicted time delays, and $\sigma_{\sdt_{ij}}$ is the 1-$\sigma$ uncertainties of the time-delay measurement.
The predicted time delays in \eref{eq:likelihood} can be expressed as
\begin{equation}
\label{eq:TDsum}
\sdt_{ij,\micro}^{\textrm{P}}=\underbrace{(\Ddt/c)\Delta\tau_{ij}}_{\textrm{cosmological time delays}}+\underbrace{t_{i,\micro}-t_{j,\micro}}_{\textrm{microlensing time delays}},
\end{equation}
where $\Delta\tau_{ij}$ is the difference of the Fermat potential at image $i$ and image $j$, and $c$ is the speed of light.
The approximation in \eref{eq:poste_new} is valid because $\macrop$ is mainly determined by $\bm{d}$, as long as there is an arc or ring due to the lensed emission of the host galaxy of the background AGN.
Because we are interested in $\Ddt$ given the microlensing model $\micro$, we can marginalize $\mt$ and $\macrop$ in \eref{eq:poste} to obtain
\begin{gather}
\begin{align}
\textrm{P}&(\Ddt|\dt,\bm{d},\micro,\macro)\nonumber\\
&=\int\int d\mt d\macrop \textrm{P}(\Ddt,\mt,\macrop|\dt,\bm{d},\micro,\macro). 
\end{align}
\end{gather}
To conservatively estimate the posterior of $\Ddt$, we should marginalize all over different microlensing models caused by different accretion disc configurations and microlensing patterns, 
\begin{gather}
\label{eq:Ddt}
\begin{align}
\textrm{P}(\Ddt|\dt,\bm{d},\macro)=&\int d\micro  \textrm{P}(\Ddt|\dt,\bm{d},\micro,\macro)\textrm{P}(\micro),
\end{align}
\end{gather}
where $\textrm{P}(\micro)$ is the prior on the configuration of the accretion disc. In this paper, we simply set a flat prior on the different configurations listed in \sref{sec:TDmap} to demonstrate this method. Thus, \eref{eq:Ddt} can be approximated as 
\begin{gather}
\begin{align}
\textrm{P}(\Ddt|\dt,\bm{d},\macro)\approx\frac{1}{N}\sum_{\micro} \textrm{P}(\Ddt|\dt,\bm{d},\micro,\macro),
\end{align}
\end{gather}
where $N$ is the number of the configurations.

\section{time-delay modeling}
\label{sec:TDmodeling}
Since the microlensing time-delay effect is an absolute, rather than fractional, error, lens systems with short time delays are expected to be affected more. Therefore, we study the impact of microlensing time-delay effect on two quadruply lensed system, \pg~as the example with short time delays in \sref{subsec:pg}, and \rxj~as the example with long time delays in \sref{subsec:rxj}.

\subsection{PG1115+080}
\label{subsec:pg}
The \pg~ source quasar, with a redshift of $\zs =1.722$, is quadruply lensed by a galaxy with $\zd = 0.31$ \citep{HenryEtal86,ChristianEtal87,Tonry98}. Among the four quasar images is an image pair $A_{1}$ and $A_{2}$ near the critical curve. As the image pair has too small a separation to be properly resolved in the seeing-limited monitoring observations, the COSMOGRAIL monitoring campaign can only obtain three light curves ($A$ light curve: the combined light curve of $A_{1}$ and $A_{2}$, $B$ light curve, and $C$ light curve; see B18 in detail), which yields two time delay measurements, $\sdt_{AC}$ and $\sdt_{BC}$\footnote{We choose $C$ as the reference image because $\sdt_{AC}$ and $\sdt_{BC}$ are the two tightest constraints. Note that the errors of there two delays are correlated.}.
We thereby need to carefully use the information from the data and prevent using the same information twice (i.e. set $\sdt_{A_{1}C}=\sdt_{A_{2}C}=\sdt_{AC}$). 

\begin{figure*}
\centering
\includegraphics[width=0.97\textwidth]{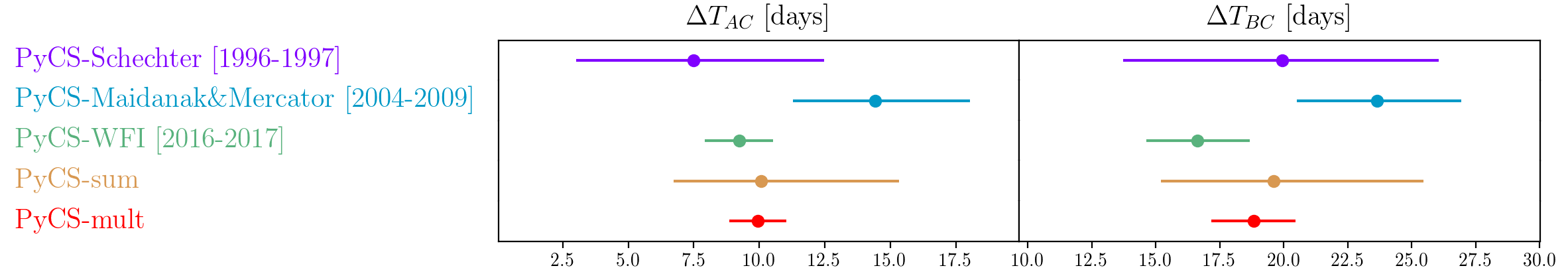}
\caption{The comparison of the time-delay measurements in different epochs by using PyCS curve-shifting algorithm (See details in B18). PyCS-Schechter [1996-1997] is computed using the Schechter data set obtained in 1996-1997, PyCS-Maidanak+Mercator [2004-2009] is computed using the Maidanak and Mercator data set in 2004-2009, PyCS-WFI [2016-2017] is computed using the WFI data set obtained in 2016-2017. PyCS-sum refers to the marginalization over the three data sets and PyCS-mult refers to the joint set of estimates. The mean values and error bars are, respectively, the 50th, 16th and 84th percentiles of the associated time-delay probability distributions.}
\label{fig:alltd}
\end{figure*}

If the difference of the $\sdt_{A_{1}C}$ and $\sdt_{A_{2}C}$ delays in the combined light curve is large enough, we can separate the measurements by doing an auto-correlation analysis on the combined light curve, which can reveal a second peak in the autocorrelation curve \citep[see e.g., Figure 3 in][]{CheungEtal14}.  If, on the other hand, the delay is too small and especially the quality of data is not good enough, the delay is indistinguishable in the combined light curve (see B18).
Therefore, the total predicted time delay between $A$ and $C$ could be approximately expressed as
\begin{equation}
\label{eq:combA}
\sdt_{AC,\micro}^{\textrm{P}}\approx\frac{F_{A_{1}}}{F_{A_{1}}+F_{A_{2}}}\sdt_{A_{1}C,\micro}^{\textrm{P}}+\frac{F_{A_{2}}}{F_{A_{1}}+F_{A_{2}}}\sdt_{A_{2}C,\micro}^{\textrm{P}},
\end{equation}
where the $F_{A_{1}}$ and $F_{A_{2}}$ are the fluxes of the $A_{1}$ and $A_{2}$ lensed quasars respectively.\footnote{The uncertainties on $F_{A_{1}}/(F_{A_{1}}+F_{A_{2}})$ and $F_{A_{2}}/(F_{A_{1}}+F_{A_{2}})$ are small enough that we can approximate them as $\approx 0$.}
Therefore, the log-likelihood of \eref{eq:likelihood} is
\begin{gather}
\label{eq:likeliPG}
\begin{align}
-&\textrm{lnP}(\dt|\Ddt,\mt,\macrop,\micro,\macro)\nonumber\\
&=\frac{(\sdt_{AC}-\sdt_{AC,\micro}^{\textrm{P}})^{2}}{2\sigma_{\sdt,AC}^{2}}+\frac{(\sdt_{BC}-\sdt_{BC,\micro}^{\textrm{P}})^{2}}{2\sigma_{\sdt,BC}^{2}}+\textrm{const},
\end{align}
\end{gather}
where ``const'' is for normalization.

\begin{figure*}
\centering
\includegraphics[width=0.95\textwidth]{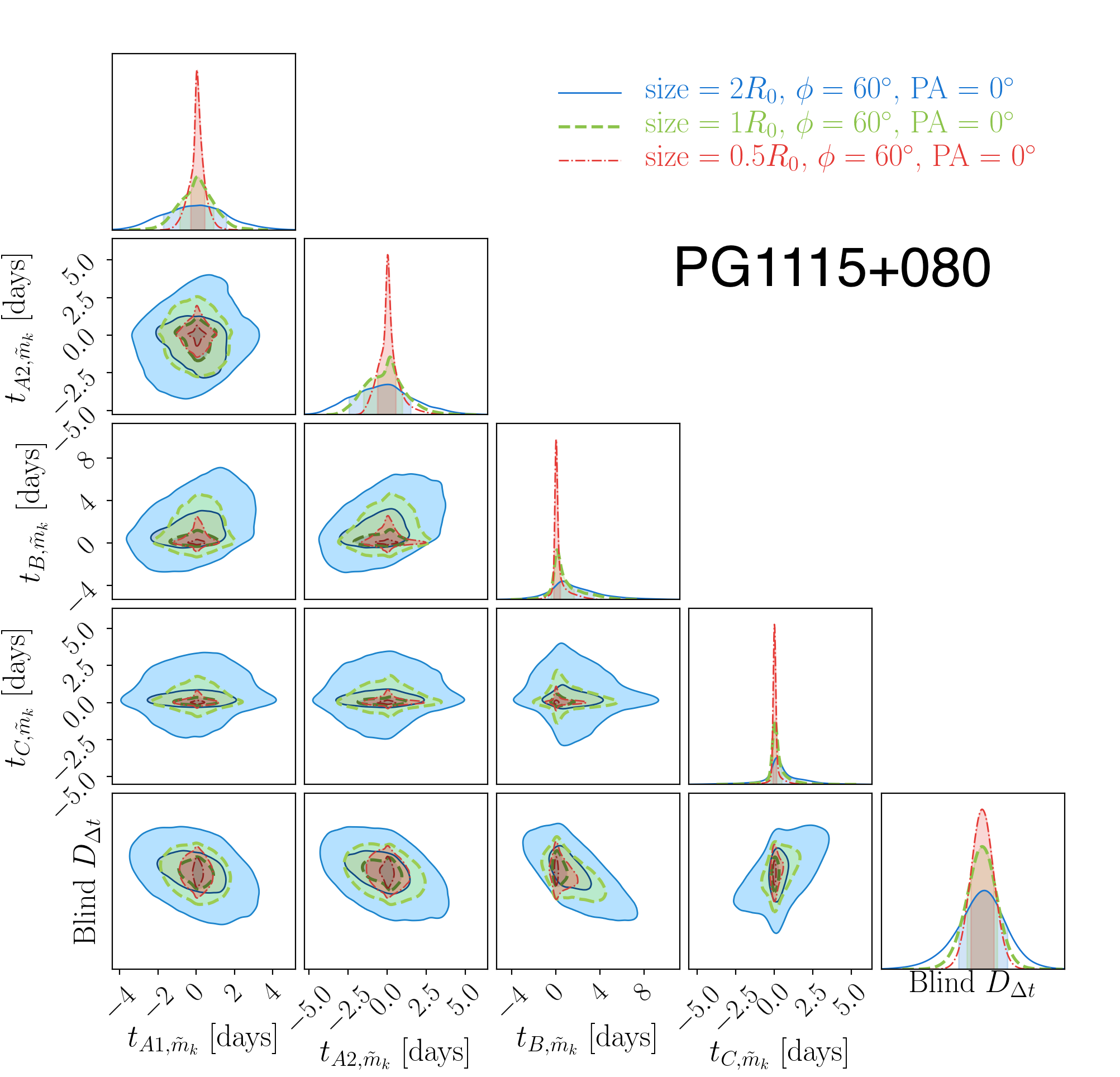}
\caption{Example showing the posteriors of the microlensing time-delay parameters at each lensed image given different sizes but the same $\phi$ and PA of the accretion disc. The results show that the smaller the disc size is, the tighter the $\Ddt$ is. We show the results with all the disc configurations we considered in \fref{fig:2TD}. The shaded regions in the marginalized 1-dimensional probability distribution functions represent the 1-sigma uncertainty.} 
\label{fig:2TD_select}
\end{figure*}
Note that we use \eref{eq:combA} and \eref{eq:likeliPG} in the analysis of a quad system with only two measured time delays. 
\eref{eq:likelihood} should be used in a more general scenario. 

B18 uses \texttt{PyCS}, a python curve-shifting toolbox containing state-of-the-art curve-shifting techniques \citep{TewesEtal13a,BonvinEtal16}, to analyze the three datasets in the different epochs (see \fref{fig:alltd}): 
\begin{itemize}
\item PyCS-Schechter: B18 use PyCS to reanalyze the dataset which was obtained with the Hiltner, WIYN, NOT and Du Pont telescopes in 1996-1997 \citep{SchechterEtal97}, 
\item PyCS-Maidanak+Mercator: B18 use PyCS to reanalyze the data which was obtained at the Maidanak telescope in 2004-2006 \citep{TsvetkovaEtal10} and Mercator telescope in 2006-2009, 
\item PyCS-WFI: B18 use PyCS to analyze the dataset which was recently obtained with ESO MPIA 2.2m telescope between December 2016 and July 2017.
\item ``PyCS-sum'' refers to the marginalization over the three data sets 
\item ``PyCS-mult'' refers to the joint set of estimates.
\end{itemize}
In \sref{sec:TD_Ddt}, we initially use PyCS-mult as input time delays and show the posterior of $\Ddt$ and $\mt$ under different source configurations as well as $\Ddt$ after marginalizing over different source configurations. 
In \sref{sec:discrepancy}, however, we argue that we should model the three time-delay measurements (PyCS-Schechter, PyCS-Maidanak+Mercator, and PyCS-WFI) simultaneously rather than use PyCS-mult. 

\begin{figure*}
\begin{minipage}{.7\linewidth}
\centering
\subfloat[]{\label{main:a}\includegraphics[scale=0.36]{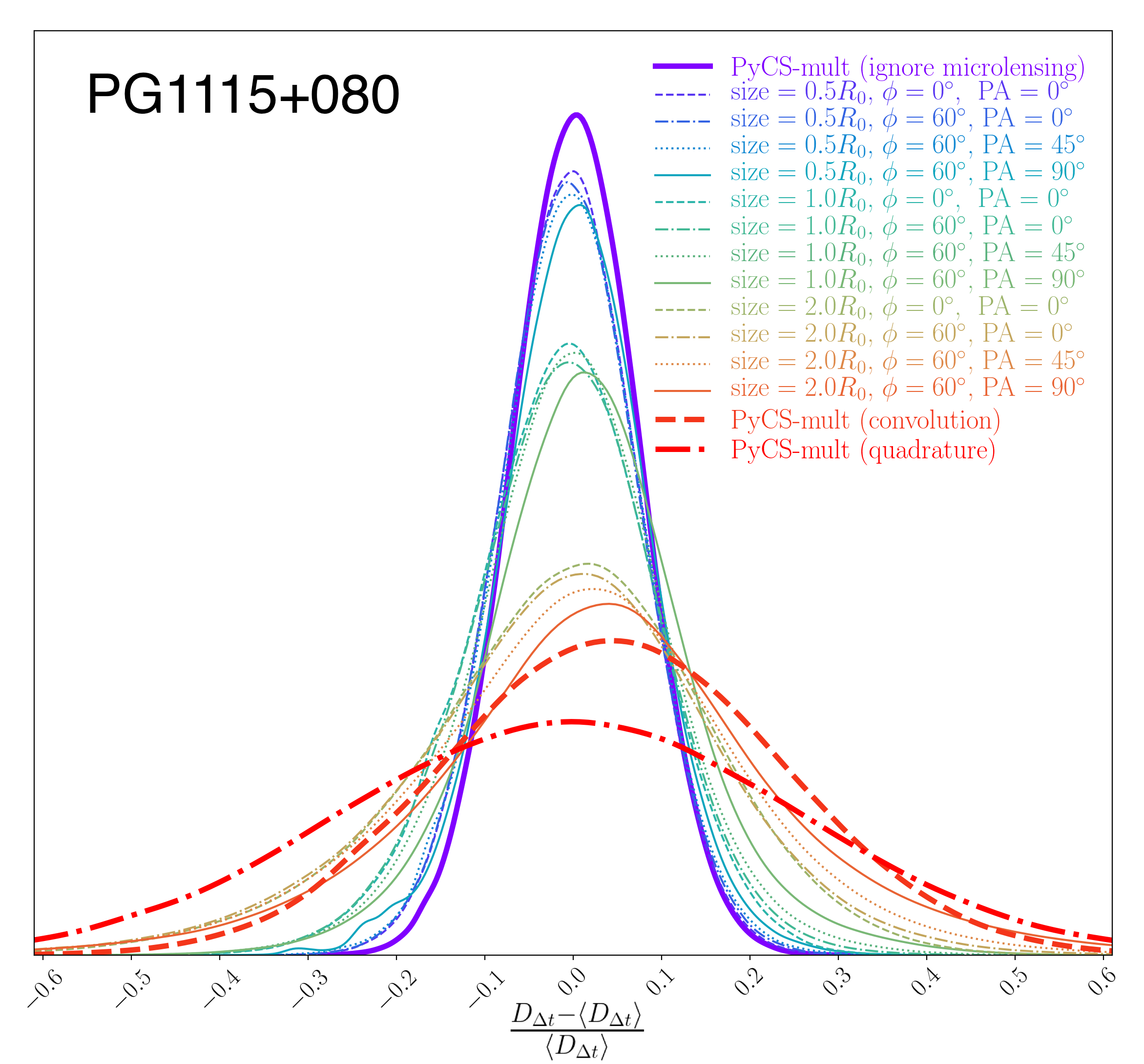}}
\end{minipage}\par\medskip
\begin{minipage}{.7\linewidth}
\centering
\subfloat[]{\label{main:b}\includegraphics[scale=0.36]{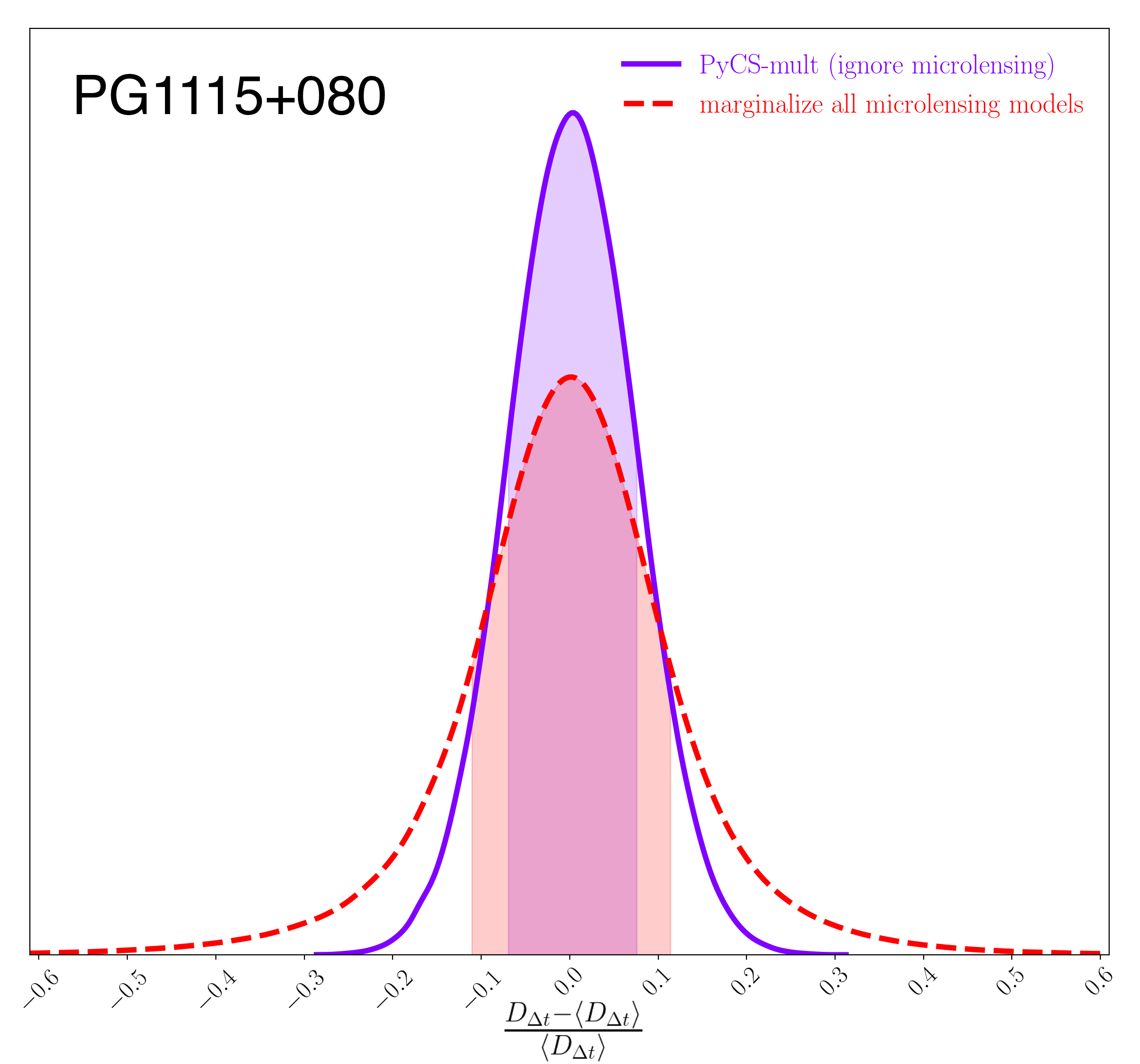}}
\end{minipage}
\centering
\caption{ The fractional difference of $\Ddt$ in different conditions. In Panel (a), the top curve shows the case which ignores the microlensing effect, the two curves in the bottom represent the cases in which we modify the PyCS-mult by convolving with the loosest case of microlensing model and by adding the loosest case of microlensing model in quadrature respectively, and the rest are the results with different accretion disc configurations. 
In Panel (b), top curve shows the case which ignores the microlensing effect (7.3\%) and the bottom curve shows the constraint on $\Ddt$ after we marginalize the different accretion discs listed in \sref{sec:TDmap} (11.3\%). The shaded regions and percentages represent the 1-sigma uncertainties.} 
\label{fig:2TD}
\end{figure*}

\subsubsection{Constraining the microlening effect and time-delay distance simultaneously}
\label{sec:TD_Ddt}
In this section, 
we use the PyCS-mult values ($\sdt_{AC}=9.9\substack{+1.1 \\ -1.1}$ days and $\sdt_{BC}=18.8\substack{+1.6 \\ -1.6}$ days) in \fref{fig:alltd} to represent the most common situation, i.e., one in which we only have a time-delay dataset from single epoch. 
In \eref{eq:poste_new}, since the $\macrop$ is dominated by the lens imaging (up to the MST), we can decouple the lens imaging modeling process and the time-delay modeling process. 
While the details of lens imaging modeling are important for measuring $H_{0}$, in this paper instead we focus on demonstrating the new time-delay modeling method developed in \sref{sec:bayes} and present the constraint on the blinded $\Ddt$\footnote{We will only unblind the results \textit{only after} coming to a consensus among the coauthors that we think we have eliminated all systematic errors, and publish the value of $H_{0}$ in Chen et al. 2018b without any modification. 
This is an important step to avoid confirmation bias \citep{Plous93}.} and the microlensing time delays.

\begin{figure*}
\centering
\includegraphics[width=0.80\textwidth]{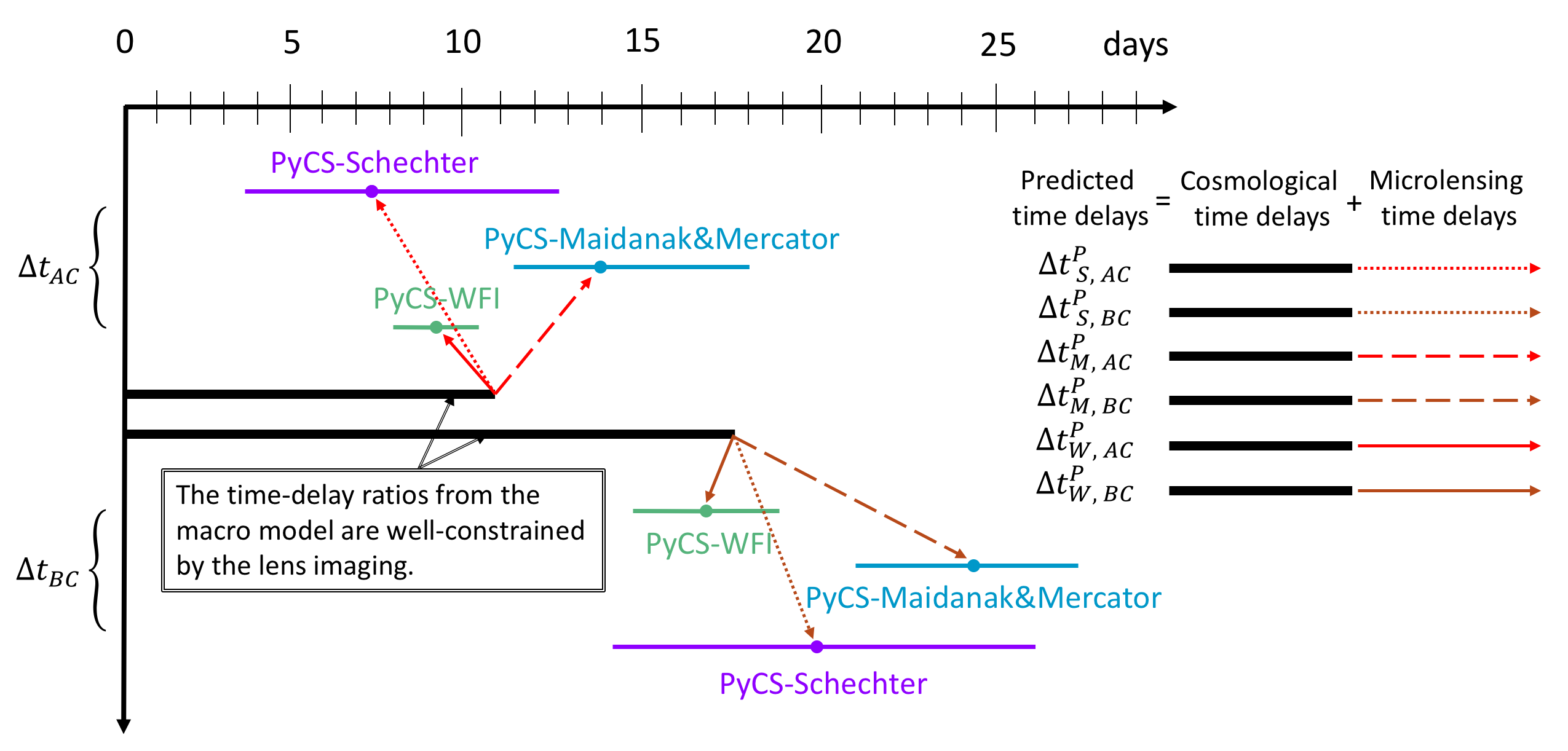}
\caption{
This figure illustrates the idea of simultaneously fitting six different time-delay measurements with single $\Ddt$ and three independent sets of microlensing parameters given the prior from the lens imaging. 
PyCS-Shechter, PyCS-Maidanak+Mercator, and PyCS-WFI represent the three time-delay measurements on $\dt_{AC}$ and three time-delay measurements on $\dt_{BC}$ in different epochs.
Since the full surface brightness of the multiple lensed images provides a strong constraint on the ratio of the Fermat potential, given a $\Ddt$ and the lens imaging, we can predict the cosmological time delays (two thick black horizontal lines). On top of the cosmological time delays, given $\mtS$, $\mtM$, $\mtW$ (i.e. the three independent microlensing parameter sets), we can obtain three sets of the predicted time-delays, $\sdt^{\textrm{P}}_{S,ij}$, $\sdt^{\textrm{P}}_{M,ij}$, $\sdt^{\textrm{P}}_{W,ij}$ (i.e. the six predicted time delays listed in the right hand side) and use the observed time delays to constrain the $\Ddt$ and the microlensing time delays. The mean values and error bars are respectively the 50th, 16th and 84th percentiles of the associated time-delay probability distributions.} 
\label{fig:method}
\end{figure*}

\fref{fig:2TD_select} shows the posteriors of the constraints on the microlensing time delays and blinded $\Ddt$ with selected accretion disc configurations. 
The most constraining case (or the case with tightest prior on microlensing time delays), i.e., with size $=0.5\textrm{R}_{0}$, $\phi =0^{\circ}$, and PA $=0^{\circ}$, provides the best constraint on $\Ddt$. 
In \fref{fig:2TD}, we show the fractional difference of $\Ddt$. 
In panel (a), the top curve represents the case which ignores the microlensing time-delay effect and the two bottom curves represent the cases in which (1) we convolve the probability distribution of the loosest constraint of microlensing (size $=2\textrm{R}_{0}$, $\phi =60^{\circ}$, and PA $=0^{\circ}$) with the probability distribution of the observed time-delays, 
(2) we simply add the uncertainty of the case with the loosest constraint on microlensing to the observed time-delay uncertainty in quadrature and shift the mean of the observed time-delay by the mean of the loosest case. 
In both cases, the constraint on $\Ddt$ are all looser than our method because both of them ignore the information from the cosmological time-delay ratios. 
The rest of the curves show the results in all different accretion disc configurations. 
Panel (a) provides two insights. First, the peaks gradually shift to larger $\Ddt$ when we increase the disc size. This makes sense as the larger the accretion disc is, the more positive the mean of the microlensing time-delay effect is (TK18). Second, the size of the accretion disc dominates the uncertainty of the inferred $\Ddt$.
Panel (b) shows the result which marginalizes all the different accretion disc configurations from panel (a).

\subsubsection{The discrepant time-delay measurements in the different epochs}
\label{sec:discrepancy}
Even though the TDC has showed that the current PyCS curve shifting technique can remove the contamination from the ``standard'' microlensing magnification effect and accurately measure time delays, PyCS-WFI and PyCS-Maidanak are $>1$ sigma discrepant.
(see PyCS-Maidanak+Mercator, and PyCS-WFI in \fref{fig:alltd})
Thus, before TK18, this raised the question of how to combine the measurements 
\begin{itemize}
\item First, we consider that we can measure the same cosmological delays on the three datasets, in which case we have three independent measurements of the delay that can be combined by multiplying their probability distribution functions. This is the PyCS-mult estimate in \fref{fig:alltd}.

\item Second, we consider that microlensing is biasing our measurements on each dataset, in which case the combined estimate is obtained by marginalizing over the three measurements because we do not have information about the microlensing time-delay effect. This is the PyCS-sum in \fref{fig:alltd}.
\end{itemize}

\begin{figure*}
  \centering
  \includegraphics[width=0.8\textwidth]{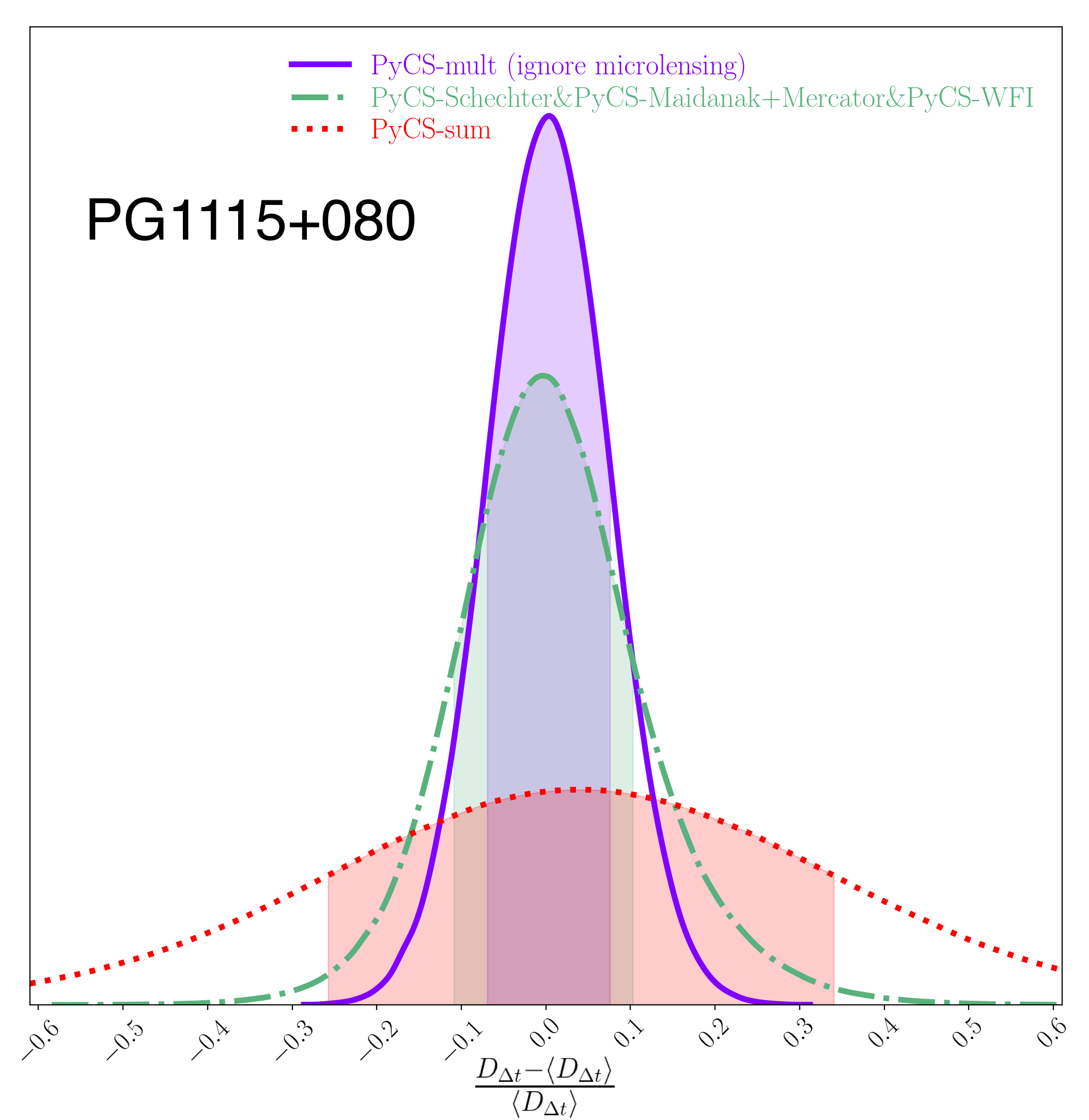}
  \caption{The comparison of the fractional difference of $\Ddt$ among the case which ignores the microlensing time-delay effect (7.3\%), the case which uses PyCS-Schechter\&PyCS-Madanak+Mercator\&PyCS-WFI (10.4\%), and the case which uses PyCS-sum (30\%). Note that the last two curves have already marginalized over all different accretion disc configurations. In the case of \pg, the uncertainty of $\Ddt$ increases from $\sim 7$\% to $\sim 10$\%.  The shaded regions and the percentages represent the 1-sigma uncertainties.}
  \label{tab:6TD_compare}
\end{figure*}

Under the assumptions of TK18, the discrepancy in the time-delay measurements in the different epochs can be understood, as the microlensing time-delay effect can vary across ten to twenty years \citep[e.g.,][]{SchechterWambsganss02,MosqueraKochanek11}. 
Therefore, we should use neither PyCS-mult nor PyCS-sum. Instead, to deliver an unbiased $\Ddt$ measurement and make good use of information from the microlensing time-delay maps and the cosmological time-delay ratios, it is better to model each time-delay measurement with its own microlensing parameter sets. 
That is, \eref{eq:poste_new} should be expanded to

\begin{gather}
\label{eq:poste_new_sixTD}
\begin{align}
\textrm{P}&(\Ddt,\mtS,\mtM,\mtW,\macrop|\dt_{S},\dt_{M},\dt_{W},\bm{d},\micro,\macro)\nonumber\\
&\approx \textrm{P}(\dt_{S},\dt_{M},\dt_{W}|\Ddt,\mtS,\mtM,\mtW,\macrop,\micro,\macro)\nonumber\\
&~~~\cdot \textrm{P}(\Ddt)\textrm{P}(\mtS,\mtM,\mtW|\micro,\macro)\textrm{P}(\macrop|\bm{d},\macro),
\end{align}
\end{gather}
where 
\begin{gather}
\label{eq:sixTDprior}
\begin{align}
\textrm{P}&(\mtS,\mtM,\mtW|\micro,\macro)\nonumber \\
&=\textrm{P}(\mtS|\micro,\macro)\textrm{P}(\mtM|\micro,\macro)\textrm{P}(\mtW|\micro,\macro)\nonumber\\
&=\prod_{i}^{N_{\textrm{im}}}\textrm{P}(t_{S,i,m_{k}}|\micro,\macro)\prod_{i}^{N_{\textrm{im}}}\textrm{P}(t_{M,i,m_{k}}|\micro,\macro)\nonumber\\ 
&~~~~~~\cdot\prod_{i}^{N_{\textrm{im}}}\textrm{P}(t_{W,i,m_{k}}|\micro,\macro),
\end{align}
\end{gather}
and the likelihood is 
\begin{gather}
\label{eq:sixTDlikelidetail}
\begin{align}
\textrm{P}&(\dt_{S},\dt_{M},\dt_{W}|\Ddt,\mtS,\mtM,\mtW,\macrop,\micro,\macro)\nonumber\\
&=\prod_{i,i<j}^{N_{\textrm{im}}}\frac{1}{\sqrt[]{2\pi}\sigma_{\sdt_{S,ij}}}\textrm{exp}\left[-\frac{(\sdt_{S,ij}-\sdt_{S,ij,\micro}^{\textrm{P}})^{2}}{2\sigma_{\sdt_{S,ij}}^{2}}\right]\nonumber\\
&~~~~\cdot\prod_{i,i<j}^{N_{\textrm{im}}}\frac{1}{\sqrt[]{2\pi}\sigma_{\sdt_{M,ij}}}\textrm{exp}\left[-\frac{(\sdt_{M,ij}-\sdt_{M,ij,\micro}^{\textrm{P}})^{2}}{2\sigma_{\sdt_{M,ij}}^{2}}\right]\nonumber\\
&~~~~~~~\cdot\prod_{i,i<j}^{N_{\textrm{im}}}\frac{1}{\sqrt[]{2\pi}\sigma_{\sdt_{W,ij}}}\textrm{exp}\left[-\frac{(\sdt_{W,ij}-\sdt_{W,ij,\micro}^{\textrm{P}})^{2}}{2\sigma_{\sdt_{W,ij}}^{2}}\right],
\end{align}
\end{gather}
where the subscript ``$S$'', ``$M$'', and ``$W$'' represents the time-delay measurements from PyCS-Schechter, PyCS-Maidanak+Mercator, and PyCS-WFI, respectively. 
\begin{table*}
    \centering
    \caption{The posteriors of the microlensing time delays at each lensed image in different datasets of \pg. The subscripts $S$, $M$, and $W$ represent the results from Schechter dataset, Maidanak+Mercator dataset, and WFI dataset, respectively. We have marginalized all the accretion disc configurations listed in \sref{sec:TDmap}. The mean values and error bars are respectively the 50th, 16th and 84th percentiles.}
    \label{tab:6TD}
    \begin{tabular}{ccccc}
        \hline
        \hline
        parameters&$t_{S,A_{1}}$ & $t_{S,A_{2}}$ & $t_{S,B}$ & $t_{S,C}~$\\
        \hline
        time delays [days]&$0.1^{+1.1}_{-1.0}$ & $0.1\pm 1.7$ & $0.06^{+1.34}_{-0.94}$ & $0.06^{+0.47}_{-0.42}$ \\ 
        \hline
        \hline
        parameters&$t_{M,A_{1}}$ & $t_{M,A_{2}}$ & $t_{M,B}$ & $t_{M,C}$ \\
        \hline
        time delays [days]&$0.08^{+1.26}_{-0.88}$ & $0.1^{+2.4}_{-1.6}$ & $0.12^{+2.45}_{-0.86}$ & $0.02^{+0.49}_{-0.41}$ \\
        \hline
        \hline
		parameters& $t_{W,A_{1}}$ & $t_{W,A_{2}}$ & $t_{W,B}$ & $t_{W,C}$ \\
        \hline
        time delays [days]& $0.07^{+0.82}_{-1.03}$ & $0.06^{+0.88}_{-1.59}$ & $0.06^{+0.94}_{-0.77}$ & $0.07^{+0.66}_{-0.43}$ \\
        \hline
		\hline
    \end{tabular}
\end{table*}
\eref{eq:sixTDlikelidetail} means that we have six measurements (two for each dataset) to constrain one $\Ddt$ and three sets of independent microlensing parameters (see \fref{fig:method}).
We assume the three datasets share the same accretion disk configuration, $\micro$, because the configuration of the accretion disc should stay invariant over the twenty years. We also follow \eref{eq:Ddt} to marginalize all the different source configurations and show the results in \tref{tab:6TD} and \fref{tab:6TD_compare}. 
\tref{tab:6TD} shows that inferred probability distribution of the microlensing time delay parameters at the position of each lensed image in different datasets. \fref{tab:6TD_compare} shows the different $\Ddt$ values when we adopt PyCS-sum and ``PyCS-Schechter\&PyCS-Madanak+Mercator\&PyCS-WFI''. Note that ``PyCS-Schechter\&PyCS-Madanak+Mercator\&PyCS-WFI'' indicates that we use three different microlensing parameter sets to model three different time delay measurements.

\begin{figure*}
\begin{minipage}{.7\linewidth}
\centering
\subfloat[]{\label{rxjmain:a}\includegraphics[scale=0.37]{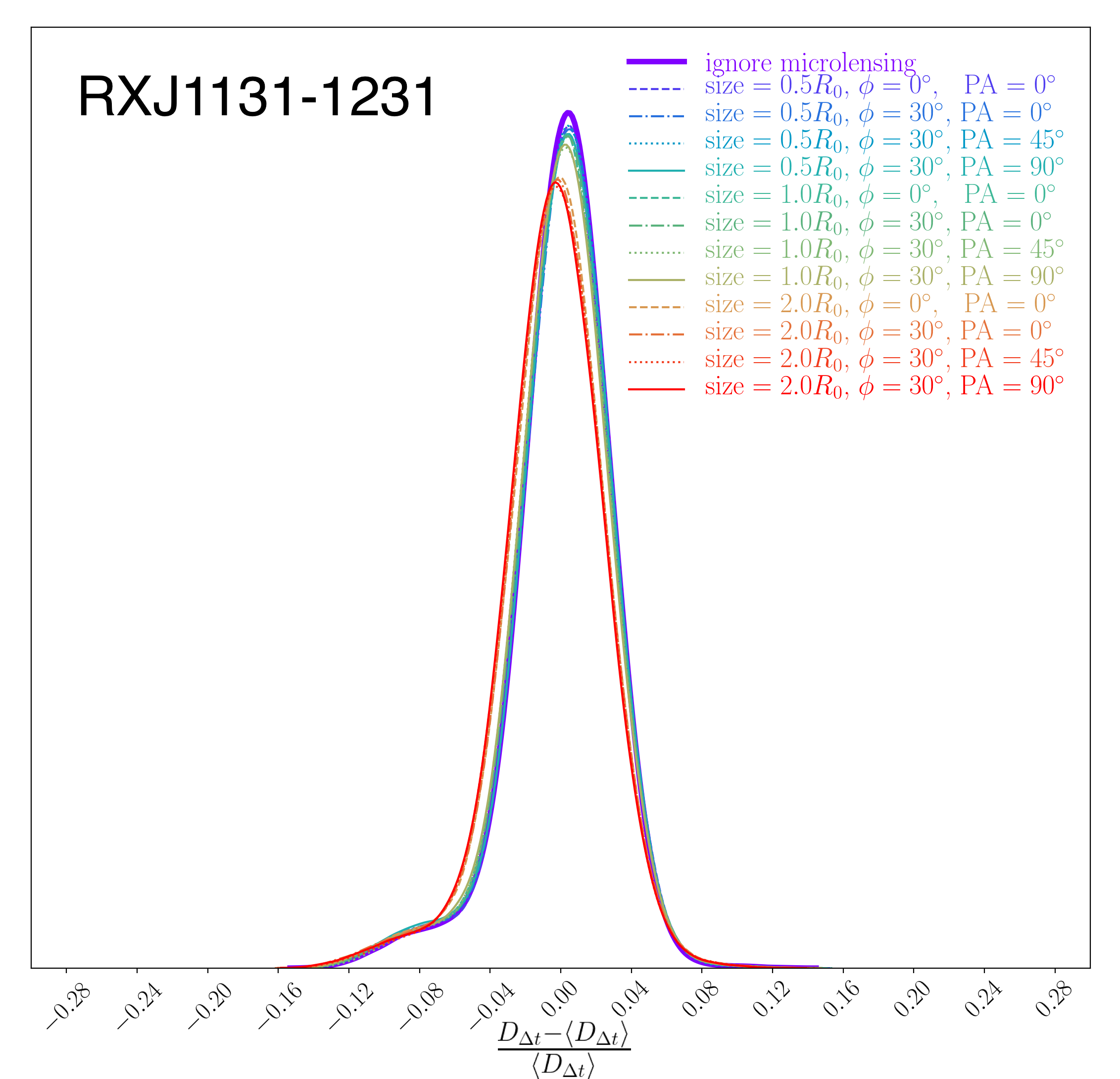}}
\end{minipage}\par\medskip
\begin{minipage}{.7\linewidth}
\centering
\subfloat[]{\label{rxjmain:b}\includegraphics[scale=0.37]{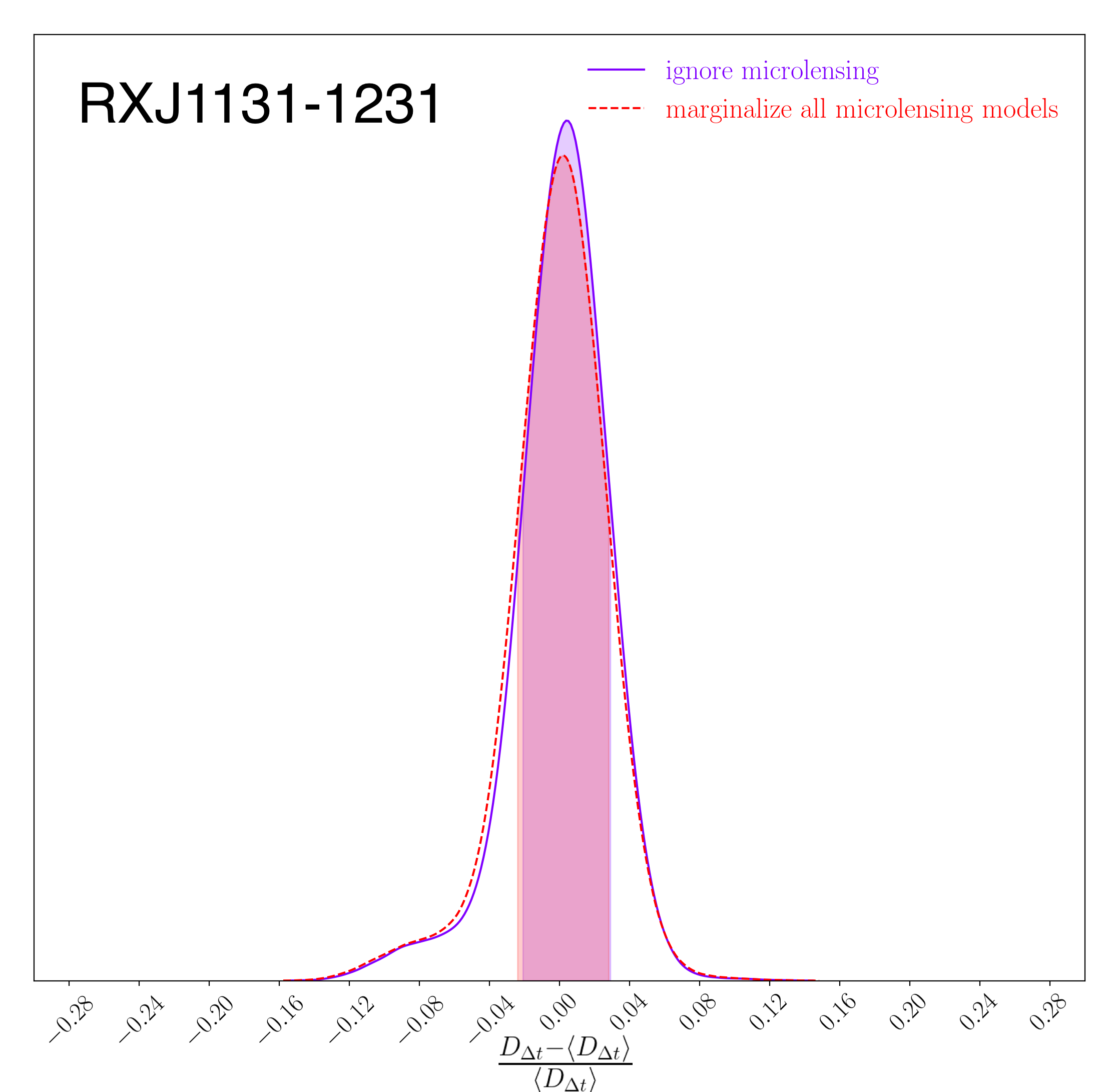}}
\end{minipage}
\centering
\caption{We present the fractional difference of $\Ddt$ in different conditions for \rxj. In Panel (a), the top curve shows the case which ignores microlensing effect and the rest are the results with different accretion disc configurations. 
In Panel (b), top curve shows the case which ignores the microlensing effect (2.5\%) and the bottom curve shows the constraint on $\Ddt$ after we marginalize the different accretion discs mentioned in \sref{subsec:rxj} (2.6\%). In the case of \rxj, the microlensing time-delay effect is negligible. The shaded regions and percentages represent the 1-sigma uncertainties.} 
\label{fig:rxj}
\end{figure*}

\begin{table}
    \centering
    \caption{The posteriors of the microlensing time delays at each lensed image of \rxj. We have marginalized all the accretion disc configurations listed in \sref{subsec:rxj}. The mean values and error bars are respectively the 50th, 16th and 84th percentiles.}
    \label{tab:rxj_micro}
    \begin{tabular}{ccccc}
        \hline
		parameters &$t_{A}$ & $t_{B}$ & $t_{C}$ & $t_{D}$\\ 
		\hline
		time delays [days] & $0.02^{+0.40}_{-0.59}$ & $0.02^{+0.38}_{-0.20}$ & $0.02^{+0.28}_{-0.21}$ & $0.01^{+0.15}_{-0.12}$ \\ 
		\hline
    \end{tabular}
\end{table}

\subsection{RXJ1131-1231}
\label{subsec:rxj}
The \rxj~system is a quadruply-lensed quasar discovered by \citet{SluseEtal03}, who also measured the spectroscopic redshifts of lensing galaxy and the background source to be at $\zd=0.295$ and $\zs=0.658$. Because of the long time delays ($\sim 90.5$ days) of this lens system, \citet{TewesEtal13a} can measure the time delay of image $D$, with a fractional uncertainty of 1.5\% ($1 \sigma$) while the delays between the three close images $A$, $B$, and $C$ are compatible with being 0 days (i.e., $\Delta_{BA}=0.5\pm1.5$ days, $\Delta_{CA}=-0.5\pm1.5$ days, $\Delta_{DA}=90.5\pm1.5$ days). Therefore, for \rxj, \eref{eq:likelihood} can be expressed as
\begin{gather}
\label{eq:TD_rxj}
\begin{align}
\textrm{P}&(\dt|\Ddt,\mt,\macrop,\micro,\macro)\nonumber\\
&=\frac{1}{\sqrt[]{2\pi}\sigma_{\sdt_{BA}}}\textrm{exp}\left[-\frac{(\sdt_{BA}-\sdt_{BA,\micro}^{\textrm{P}})^{2}}{2\sigma_{\sdt_{BA}}^{2}}\right]\nonumber\\
&~~~~~~~~\cdot\frac{1}{\sqrt[]{2\pi}\sigma_{\sdt_{CA}}}\textrm{exp}\left[-\frac{(\sdt_{CA}-\sdt_{CA,\micro}^{\textrm{P}})^{2}}{2\sigma_{\sdt_{CA}}^{2}}\right]\nonumber\\
&~~~~~~~~~~~~~~\cdot\frac{1}{\sqrt[]{2\pi}\sigma_{\sdt_{DA}}}\textrm{exp}\left[-\frac{(\sdt_{DA}-\sdt_{DA,\micro}^{\textrm{P}})^{2}}{2\sigma_{\sdt_{DA}}^{2}}\right].
\end{align}
\end{gather}
We use the same $\kappa$, $\gamma$, and $\kappa_{\star}/\kappa$ as TK18 to generate the microlensing time-delay maps given the combinations of different accretion disc sizes ($0.5R_{0}$, $1R_{0}$, and $2R_{0}$), different $\phi$ to the line of slight ($0^{\circ}$ and $30^{\circ}$), and different PA ($0^{\circ}$, $45^{\circ}$, and $90^{\circ}$) at the four lensed images. We show the constraint on $\Ddt$ in different accretion disc configurations in \fref{fig:rxj} and the marginalized posteriors of the microlensing time delays in \tref{tab:rxj_micro}.
As expected, the microlensing time-delay effect on the lens with longer time delays has less impact. In the case of \rxj, the impact by microlensing time-delay effect is negligible.

\section{Conclusions}
\label{sec:conclusion}
This paper quantifies the impact of microlensing time delays, produced under the assumption that AGN variability is the lamp-post type, on the time-delay distance. For that purpose we calculate the time-delay distance, $\Ddt$, including the microlensing time-delay effect for two lens systems, \pg~and \rxj, We find that this broadens the probability distribution by about 3\% in the case of \pg~and 0.1\% in the case of \rxj.

Given the lamp-post model assumption, although we do not have any knowledge about how severely each light curve is affected by the microlensing time-delay effect, the cosmological time-delay ratios, which are well-constrained by the full surface brightness morphology of the lensed host galaxy emission, provide the constraining information on the possible combinations of the microlensing time delay at each lensed image position. 
Furthermore, the microlensing time-delay maps also provide constraints on the microlensing time-delay effect at each lensed image position. Thus,
we have developed a new time-delay prediction model, which uses the information from cosmological time-delay ratios, as well as the information from microlensing time-delay maps, to remove the biases caused by this proposed effect under a Bayesian framework. 

There are several key results: 
\begin{enumerate}
\item Under the assumption of TK18, different lens systems can yield different $H_{0}$ due to the fact that the measured time delays are not the cosmological time delays but the combination of cosmological time delays plus microlensing time delays.
With this new time-delay prediction model, we can separately predict the cosmological time delays and microlensing time delays to measure a unbiased value of $H_{0}$ for each lens. 
Thus, this paper addresses concerns that TDSL have already hit the systematics floor in the time delay measurements due to this newly proposed microlensing time-delay effect, although it does increase the error budget.
\item The time-delay measurements in different epochs should be modeled by different microlensing parameters as they are likely affected by different microlensing time-delay effects.
\item Given a lens system, the constraint on $\Ddt$ mainly depends on the size of the accretion disc, whereas the inclination and the position angle of the disc have little influence. Thus, the smaller the disc is, the smaller the variances on the microlensing time delays are. 
\item The uncertainty on $\Ddt$ from \pg, which has relatively short time delays, increases from $\sim 7$\% to $\sim 10$\% when we include the microlensing time-delay effects. Without our new technique, the uncertainty on $\Ddt$ from \pg~can increase by up to 30\%.
\item The uncertainty on $\Ddt$ from \rxj, which has relatively long time delays, increases only from $\sim 2.5$\% to $\sim 2.6$\% when we include the microlensing time-delay effects. Thus, the impact of the microlensing time-delay effect on \rxj~is negligible.
\end{enumerate}
Note that although we assume the lamp-post model on accretion disc, there is evidence \citep[e.g.,][]{MorganEtal10,BlackburneEtal11} showing that the size of the accretion disc is larger than the prediction from the standard thin disk theory. In addition, exist a variety of alternative accretion disc models \citep[e.g.,][]{Beloborodov99}, including e.g. the inhomogeneous accretion disc \citep[e.g.,][]{DexterAgol11} for which variability is different from the lamp-post model.

As the amplitude of this effect highly depends on the accretion disk models, which are not well-understood, in our future determinations of $H_{0}$ from the  H0LiCOW programme, we will present the measurements both with and without microlensing time-delay effects. 
The techniques for verifying the accretion disk models by using observational data are currently under development.

Finally, we want to stress that with the advantage of cosmological time-delay ratios, quads are better than doubles in term of constraining the microlensing time-delay effect in measuring the value of $H_{0}$. The final $H_{0}$ measurements from \pg~and \rxj~will be presented in Chen et al. 2018b in prep. 

\section*{Acknowledgements}
We thank Cristian E. Rusu and Stefan Hilbert for comments. G.~C.-F.~C. also thanks Chun-Hao To and Shih-Wei Chuo for  technical discussions, and the UC Davis cosmology group for providing a friendly research environment. G.~C.-F.~C. acknowledges support from the Ministry of Education in Taiwan via Government Scholarship to Study Abroad (GSSA). G.~C.-F.~C. and C. D. F. acknowledge support from the National Science Foundation under grant AST-1715611.  G.~C.-F.~C. and C. D. F. thank the Max Planck Institute for Astrophysics for kind hospitality during working visits.  K.C.W. is supported by an EACOA Fellowship awarded by the East Asia Core Observatories Association, which consists of the Academia Sinica Institute of Astronomy and Astrophysics, the National Astronomical Observatory of Japan, the National Astronomical Observatories of the Chinese Academy of Sciences, and the Korea Astronomy and Space Science Institute. J.~C., V.~B., M.~M., and F.~C. are supported by the Swiss National Science Foundation. S.~V. has received funding from the European Research Council (ERC) under the European Union's Horizon 2020 research and innovation programme (grant agreement No 758853). L.~V.~E.~K. are supported through an NWO-VICI grant (project number 639.043.308). S.~H.~S thanks the Max Planck Society for support through the Max Planck Research Group. A.~J.~S. and T.~T. acknowledge support by NASA through STSCI grant HST-GO-15320, and by the Packard Foundation through a Packard Fellowship to T.~T.




\bibliographystyle{mnras}
\bibliography{AO_cosmography} 

\begin{thebibliography}{}
\makeatletter
\relax
\def\mn@urlcharsother{\let\do\@makeother \do\$\do\&\do\#\do\^\do\_\do\%\do\~}
\def\mn@doi{\begingroup\mn@urlcharsother \@ifnextchar [ {\mn@doi@}
  {\mn@doi@[]}}
\def\mn@doi@[#1]#2{\def\@tempa{#1}\ifx\@tempa\@empty \href
  {http://dx.doi.org/#2} {doi:#2}\else \href {http://dx.doi.org/#2} {#1}\fi
  \endgroup}
\def\mn@eprint#1#2{\mn@eprint@#1:#2::\@nil}
\def\mn@eprint@arXiv#1{\href {http://arxiv.org/abs/#1} {{\tt arXiv:#1}}}
\def\mn@eprint@dblp#1{\href {http://dblp.uni-trier.de/rec/bibtex/#1.xml}
  {dblp:#1}}
\def\mn@eprint@#1:#2:#3:#4\@nil{\def\@tempa {#1}\def\@tempb {#2}\def\@tempc
  {#3}\ifx \@tempc \@empty \let \@tempc \@tempb \let \@tempb \@tempa \fi \ifx
  \@tempb \@empty \def\@tempb {arXiv}\fi \@ifundefined
  {mn@eprint@\@tempb}{\@tempb:\@tempc}{\expandafter \expandafter \csname
  mn@eprint@\@tempb\endcsname \expandafter{\@tempc}}}

\bibitem[\protect\citeauthoryear{{Agnello}}{{Agnello}}{2017}]{Agnello17}
{Agnello} A.,  2017, preprint, \href
  {http://adsabs.harvard.edu/abs/2017arXiv170508900A} {} (\mn@eprint {arXiv}
  {1705.08900})

\bibitem[\protect\citeauthoryear{{Agnello} et~al.,}{{Agnello}
  et~al.}{2017}]{AgnelloEtal17}
{Agnello} A.,  et~al., 2017, preprint, \href
  {http://adsabs.harvard.edu/abs/2017arXiv171103971A} {} (\mn@eprint {arXiv}
  {1711.03971})

\bibitem[\protect\citeauthoryear{{Aubourg} et~al.,}{{Aubourg}
  et~al.}{2015}]{AuborugEtal15}
{Aubourg} {\'E}.,  et~al., 2015, \mn@doi [\prd] {10.1103/PhysRevD.92.123516},
  \href {http://adsabs.harvard.edu/abs/2015PhRvD..92l3516A} {92, 123516}

\bibitem[\protect\citeauthoryear{{Avestruz}, {Li}, {Lightman}, {Collett}  \&
  {Luo}}{{Avestruz} et~al.}{2017}]{AvestruzEtal17}
{Avestruz} C.,  {Li} N.,  {Lightman} M.,  {Collett} T.~E.,   {Luo} W.,  2017,
  preprint, \href {http://adsabs.harvard.edu/abs/2017arXiv170402322A} {}
  (\mn@eprint {arXiv} {1704.02322})

\bibitem[\protect\citeauthoryear{{Beloborodov}}{{Beloborodov}}{1999}]{Beloborodov99}
{Beloborodov} A.~M.,  1999, in {Poutanen} J.,  {Svensson} R.,  eds,
  Astronomical Society of the Pacific Conference Series Vol. 161, High Energy
  Processes in Accreting Black Holes. p.~295 (\mn@eprint {} {astro-ph/9901108})

\bibitem[\protect\citeauthoryear{{Birrer}, {Amara}  \& {Refregier}}{{Birrer}
  et~al.}{2015}]{BirrerEtal15}
{Birrer} S.,  {Amara} A.,   {Refregier} A.,  2015, \mn@doi [\apj]
  {10.1088/0004-637X/813/2/102}, \href
  {http://adsabs.harvard.edu/abs/2015ApJ...813..102B} {813, 102}

\bibitem[\protect\citeauthoryear{{Birrer}, {Welschen}, {Amara}  \&
  {Refregier}}{{Birrer} et~al.}{2017}]{BirrerEtal17_LOS}
{Birrer} S.,  {Welschen} C.,  {Amara} A.,   {Refregier} A.,  2017, \mn@doi
  [\jcap] {10.1088/1475-7516/2017/04/049}, \href
  {http://adsabs.harvard.edu/abs/2017JCAP...04..049B} {4, 049}

\bibitem[\protect\citeauthoryear{{Blackburne}, {Pooley}, {Rappaport}  \&
  {Schechter}}{{Blackburne} et~al.}{2011}]{BlackburneEtal11}
{Blackburne} J.~A.,  {Pooley} D.,  {Rappaport} S.,   {Schechter} P.~L.,  2011,
  \mn@doi [\apj] {10.1088/0004-637X/729/1/34}, \href
  {http://adsabs.harvard.edu/abs/2011ApJ...729...34B} {729, 34}

\bibitem[\protect\citeauthoryear{{Bonvin}, {Tewes}, {Courbin}, {Kuntzer},
  {Sluse}  \& {Meylan}}{{Bonvin} et~al.}{2016}]{BonvinEtal16}
{Bonvin} V.,  {Tewes} M.,  {Courbin} F.,  {Kuntzer} T.,  {Sluse} D.,   {Meylan}
  G.,  2016, \mn@doi [\aap] {10.1051/0004-6361/201526704}, \href
  {http://adsabs.harvard.edu/abs/2016A%26A...585A..88B} {585, A88}

\bibitem[\protect\citeauthoryear{{Bonvin} et~al.,}{{Bonvin}
  et~al.}{2017}]{BonvinEtal17}
{Bonvin} V.,  et~al., 2017, \mn@doi [\mnras] {10.1093/mnras/stw3006}, \href
  {http://adsabs.harvard.edu/abs/2017MNRAS.465.4914B} {465, 4914}

\bibitem[\protect\citeauthoryear{{Chen} et~al.,}{{Chen}
  et~al.}{2016}]{GChenEtal16}
{Chen} G.~C.-F.,  et~al., 2016, \mn@doi [\mnras] {10.1093/mnras/stw991}, \href
  {http://adsabs.harvard.edu/abs/2016MNRAS.462.3457C} {462, 3457}

\bibitem[\protect\citeauthoryear{{Cheung} et~al.,}{{Cheung}
  et~al.}{2014}]{CheungEtal14}
{Cheung} C.~C.,  et~al., 2014, \mn@doi [\apjl] {10.1088/2041-8205/782/2/L14},
  \href {http://adsabs.harvard.edu/abs/2014ApJ...782L..14C} {782, L14}

\bibitem[\protect\citeauthoryear{{Christian}, {Crabtree}  \&
  {Waddell}}{{Christian} et~al.}{1987}]{ChristianEtal87}
{Christian} C.~A.,  {Crabtree} D.,   {Waddell} P.,  1987, \mn@doi [\apj]
  {10.1086/164847}, \href {http://adsabs.harvard.edu/abs/1987ApJ...312...45C}
  {312, 45}

\bibitem[\protect\citeauthoryear{{Courbin} et~al.,}{{Courbin}
  et~al.}{2018}]{CourbinEtal18}
{Courbin} F.,  et~al., 2018, \mn@doi [\aap] {10.1051/0004-6361/201731461},
  \href {http://adsabs.harvard.edu/abs/2018A%26A...609A..71C} {609, A71}

\bibitem[\protect\citeauthoryear{{Dexter} \& {Agol}}{{Dexter} \&
  {Agol}}{2011}]{DexterAgol11}
{Dexter} J.,  {Agol} E.,  2011, \mn@doi [\apjl] {10.1088/2041-8205/727/1/L24},
  \href {http://adsabs.harvard.edu/abs/2011ApJ...727L..24D} {727, L24}

\bibitem[\protect\citeauthoryear{{Ding} et~al.,}{{Ding}
  et~al.}{2018}]{DingEtal18}
{Ding} X.,  et~al., 2018, preprint, \href
  {http://adsabs.harvard.edu/abs/2018arXiv180101506D} {} (\mn@eprint {arXiv}
  {1801.01506})

\bibitem[\protect\citeauthoryear{{Dobler}, {Fassnacht}, {Treu}, {Marshall},
  {Liao}, {Hojjati}, {Linder}  \& {Rumbaugh}}{{Dobler}
  et~al.}{2013}]{DoblerEtal13}
{Dobler} G.,  {Fassnacht} C.,  {Treu} T.,  {Marshall} P.~J.,  {Liao} K.,
  {Hojjati} A.,  {Linder} E.,   {Rumbaugh} N.,  2013, preprint, \href
  {http://adsabs.harvard.edu/abs/2013arXiv1310.4830D} {} (\mn@eprint {arXiv}
  {1310.4830})

\bibitem[\protect\citeauthoryear{{Eulaers} et~al.,}{{Eulaers}
  et~al.}{2013}]{EulaersEtal13}
{Eulaers} E.,  et~al., 2013, \mn@doi [\aap] {10.1051/0004-6361/201321140},
  \href {http://adsabs.harvard.edu/abs/2013A%26A...553A.121E} {553, A121}

\bibitem[\protect\citeauthoryear{{Falco}, {Gorenstein}  \& {Shapiro}}{{Falco}
  et~al.}{1985}]{FalcoEtal85}
{Falco} E.~E.,  {Gorenstein} M.~V.,   {Shapiro} I.~I.,  1985, \mn@doi [\apjl]
  {10.1086/184422}, \href {http://adsabs.harvard.edu/abs/1985ApJ...289L...1F}
  {289, L1}

\bibitem[\protect\citeauthoryear{{Fassnacht}, {Xanthopoulos}, {Koopmans}  \&
  {Rusin}}{{Fassnacht} et~al.}{2002}]{FassnachtEtal02}
{Fassnacht} C.~D.,  {Xanthopoulos} E.,  {Koopmans} L.~V.~E.,   {Rusin} D.,
  2002, \mn@doi [\apj] {10.1086/344368}, \href
  {http://adsabs.harvard.edu/cgi-bin/nph-bib_query?bibcode=2002ApJ...5
  81..823F&db_key=AST} {581, 823}

\bibitem[\protect\citeauthoryear{{Fassnacht}, {Koopmans}  \&
  {Wong}}{{Fassnacht} et~al.}{2011}]{FassnachtEtal11}
{Fassnacht} C.~D.,  {Koopmans} L.~V.~E.,   {Wong} K.~C.,  2011, \mn@doi
  [\mnras] {10.1111/j.1365-2966.2010.17591.x}, \href
  {http://adsabs.harvard.edu/abs/2011MNRAS.410.2167F} {410, 2167}

\bibitem[\protect\citeauthoryear{{Foreman-Mackey}, {Hogg}, {Lang}  \&
  {Goodman}}{{Foreman-Mackey} et~al.}{2013}]{Foreman-MackeyEtal13}
{Foreman-Mackey} D.,  {Hogg} D.~W.,  {Lang} D.,   {Goodman} J.,  2013, \mn@doi
  [\pasp] {10.1086/670067}, \href
  {http://adsabs.harvard.edu/abs/2013PASP..125..306F} {125, 306}

\bibitem[\protect\citeauthoryear{{Freedman}}{{Freedman}}{2017}]{Freedman17}
{Freedman} W.~L.,  2017, \mn@doi [Nature Astronomy] {10.1038/s41550-017-0169},
  \href {http://adsabs.harvard.edu/abs/2017NatAs...1E.169F} {1, 0169}

\bibitem[\protect\citeauthoryear{{Goodman} \& {Weare}}{{Goodman} \&
  {Weare}}{2010}]{GoodmanWeare10}
{Goodman} J.,  {Weare} J.,  2010, \mn@doi [Communications in Applied
  Mathematics and Computational Science, Vol.~5, No.~1, p.~65-80, 2010]
  {10.2140/camcos.2010.5.65}, \href
  {http://adsabs.harvard.edu/abs/2010CAMCS...5...65G} {5, 65}

\bibitem[\protect\citeauthoryear{{Henry} \& {Heasley}}{{Henry} \&
  {Heasley}}{1986}]{HenryEtal86}
{Henry} J.~P.,  {Heasley} J.~N.,  1986, \mn@doi [\nat] {10.1038/321139a0},
  \href {http://adsabs.harvard.edu/abs/1986Natur.321..139H} {321, 139}

\bibitem[\protect\citeauthoryear{{Hinshaw} et~al.,}{{Hinshaw}
  et~al.}{2013}]{HinshawEtal13}
{Hinshaw} G.,  et~al., 2013, \mn@doi [\apjs] {10.1088/0067-0049/208/2/19},
  \href {http://adsabs.harvard.edu/abs/2013ApJS..208...19H} {208, 19}

\bibitem[\protect\citeauthoryear{{Hinton}}{{Hinton}}{2016}]{Hinton16}
{Hinton} S.~R.,  2016, \mn@doi [The Journal of Open Source Software]
  {10.21105/joss.00045}, \href
  {http://adsabs.harvard.edu/abs/2016JOSS....1...45H} {1, 00045}

\bibitem[\protect\citeauthoryear{{Jee}, {Komatsu}  \& {Suyu}}{{Jee}
  et~al.}{2015}]{JeeEtal15}
{Jee} I.,  {Komatsu} E.,   {Suyu} S.~H.,  2015, \mn@doi [\jcap]
  {10.1088/1475-7516/2015/11/033}, \href
  {http://adsabs.harvard.edu/abs/2015JCAP...11..033J} {11, 033}

\bibitem[\protect\citeauthoryear{{Jee}, {Komatsu}, {Suyu}  \& {Huterer}}{{Jee}
  et~al.}{2016}]{JeeEtal16}
{Jee} I.,  {Komatsu} E.,  {Suyu} S.~H.,   {Huterer} D.,  2016, \mn@doi [\jcap]
  {10.1088/1475-7516/2016/04/031}, \href
  {http://adsabs.harvard.edu/abs/2016JCAP...04..031J} {4, 031}

\bibitem[\protect\citeauthoryear{{Joseph} et~al.,}{{Joseph}
  et~al.}{2014}]{JosephEtal14}
{Joseph} R.,  et~al., 2014, \mn@doi [\aap] {10.1051/0004-6361/201423365}, \href
  {http://adsabs.harvard.edu/abs/2014A%26A...566A..63J} {566, A63}

\bibitem[\protect\citeauthoryear{{Keeton} \& {Moustakas}}{{Keeton} \&
  {Moustakas}}{2009}]{KeetonEtal09}
{Keeton} C.~R.,  {Moustakas} L.~A.,  2009, \mn@doi [\apj]
  {10.1088/0004-637X/699/2/1720}, \href
  {http://adsabs.harvard.edu/abs/2009ApJ...699.1720K} {699, 1720}

\bibitem[\protect\citeauthoryear{{Kochanek} \& {Dalal}}{{Kochanek} \&
  {Dalal}}{2004}]{KochanekDalal04}
{Kochanek} C.~S.,  {Dalal} N.,  2004, \mn@doi [\apj] {10.1086/421436}, \href
  {http://adsabs.harvard.edu/abs/2004ApJ...610...69K} {610, 69}

\bibitem[\protect\citeauthoryear{{Komatsu} et~al.,}{{Komatsu}
  et~al.}{2011}]{KomatsuEtal11}
{Komatsu} E.,  et~al., 2011, \mn@doi [\apjs] {10.1088/0067-0049/192/2/18},
  \href {http://adsabs.harvard.edu/abs/2011ApJS..192...18K} {192, 18}

\bibitem[\protect\citeauthoryear{{Koopmans}}{{Koopmans}}{2005}]{Koopmans05}
{Koopmans} L.~V.~E.,  2005, \mn@doi [\mnras]
  {10.1111/j.1365-2966.2005.09523.x}, \href
  {http://adsabs.harvard.edu/cgi-bin/nph-bib_query?bibcode=2005MNRAS.363.1136K&db_key=AST}
  {363, 1136}

\bibitem[\protect\citeauthoryear{{Koopmans}, {Treu}, {Fassnacht}, {Blandford}
  \& {Surpi}}{{Koopmans} et~al.}{2003}]{KoopmansEtal03}
{Koopmans} L.~V.~E.,  {Treu} T.,  {Fassnacht} C.~D.,  {Blandford} R.~D.,
  {Surpi} G.,  2003, \mn@doi [\apj] {10.1086/379226}, \href
  {http://adsabs.harvard.edu/cgi-bin/nph-bib_query?bibcode=2003ApJ...5
  99...70K&db_key=AST} {599, 70}

\bibitem[\protect\citeauthoryear{{Lanusse}, {Ma}, {Li}, {Collett}, {Li},
  {Ravanbakhsh}, {Mandelbaum}  \& {P{\'o}czos}}{{Lanusse}
  et~al.}{2018}]{LanusseEtal18}
{Lanusse} F.,  {Ma} Q.,  {Li} N.,  {Collett} T.~E.,  {Li} C.-L.,  {Ravanbakhsh}
  S.,  {Mandelbaum} R.,   {P{\'o}czos} B.,  2018, \mn@doi [\mnras]
  {10.1093/mnras/stx1665}, \href
  {http://adsabs.harvard.edu/abs/2018MNRAS.473.3895L} {473, 3895}

\bibitem[\protect\citeauthoryear{{Liao} et~al.,}{{Liao}
  et~al.}{2015}]{LiaoEtal15}
{Liao} K.,  et~al., 2015, \mn@doi [\apj] {10.1088/0004-637X/800/1/11}, \href
  {http://adsabs.harvard.edu/abs/2015ApJ...800...11L} {800, 11}

\bibitem[\protect\citeauthoryear{{Lin} et~al.,}{{Lin} et~al.}{2017}]{LinEtal17}
{Lin} H.,  et~al., 2017, \mn@doi [\apjl] {10.3847/2041-8213/aa624e}, \href
  {http://adsabs.harvard.edu/abs/2017ApJ...838L..15L} {838, L15}

\bibitem[\protect\citeauthoryear{{Mao} \& {Schneider}}{{Mao} \&
  {Schneider}}{1998}]{MaoSchneider98}
{Mao} S.,  {Schneider} P.,  1998, \mnras, \href
  {http://adsabs.harvard.edu/abs/1998MNRAS.295..587M} {295, 587}

\bibitem[\protect\citeauthoryear{{Morgan}, {Kochanek}, {Morgan}  \&
  {Falco}}{{Morgan} et~al.}{2010}]{MorganEtal10}
{Morgan} C.~W.,  {Kochanek} C.~S.,  {Morgan} N.~D.,   {Falco} E.~E.,  2010,
  \mn@doi [\apj] {10.1088/0004-637X/712/2/1129}, \href
  {http://adsabs.harvard.edu/abs/2010ApJ...712.1129M} {712, 1129}

\bibitem[\protect\citeauthoryear{{Mosquera} \& {Kochanek}}{{Mosquera} \&
  {Kochanek}}{2011}]{MosqueraKochanek11}
{Mosquera} A.~M.,  {Kochanek} C.~S.,  2011, \mn@doi [\apj]
  {10.1088/0004-637X/738/1/96}, \href
  {http://adsabs.harvard.edu/abs/2011ApJ...738...96M} {738, 96}

\bibitem[\protect\citeauthoryear{{Nightingale} \& {Dye}}{{Nightingale} \&
  {Dye}}{2015}]{NightingaleDye15}
{Nightingale} J.~W.,  {Dye} S.,  2015, \mn@doi [\mnras]
  {10.1093/mnras/stv1455}, \href
  {http://adsabs.harvard.edu/abs/2015MNRAS.452.2940N} {452, 2940}

\bibitem[\protect\citeauthoryear{{Oguri}}{{Oguri}}{2010}]{OguriAlgorithm10}
{Oguri} M.,  2010, \mn@doi [\pasj] {10.1093/pasj/62.4.1017}, \href
  {http://adsabs.harvard.edu/abs/2010PASJ...62.1017O} {62, 1017}

\bibitem[\protect\citeauthoryear{{Oguri} \& {Marshall}}{{Oguri} \&
  {Marshall}}{2010}]{OguriMarshall10}
{Oguri} M.,  {Marshall} P.~J.,  2010, \mn@doi [\mnras]
  {10.1111/j.1365-2966.2010.16639.x}, \href
  {http://adsabs.harvard.edu/abs/2010MNRAS.405.2579O} {405, 2579}

\bibitem[\protect\citeauthoryear{{Ostrovski} et~al.,}{{Ostrovski}
  et~al.}{2017}]{OstrovskiEtal17}
{Ostrovski} F.,  et~al., 2017, \mn@doi [\mnras] {10.1093/mnras/stw2958}, \href
  {http://adsabs.harvard.edu/abs/2017MNRAS.465.4325O} {465, 4325}

\bibitem[\protect\citeauthoryear{{Ostrovski} et~al.,}{{Ostrovski}
  et~al.}{2018}]{OstrovskiEtal18}
{Ostrovski} F.,  et~al., 2018, \mn@doi [\mnras] {10.1093/mnrasl/slx173}, \href
  {http://adsabs.harvard.edu/abs/2018MNRAS.473L.116O} {473, L116}

\bibitem[\protect\citeauthoryear{{Peng}, {Impey}, {Rix}, {Falco}, {Keeton},
  {Kochanek}, {Leh{\'a}r}  \& {McLeod}}{{Peng} et~al.}{2006}]{PengEtal06}
{Peng} C.~Y.,  {Impey} C.~D.,  {Rix} H.-W.,  {Falco} E.~E.,  {Keeton} C.~R.,
  {Kochanek} C.~S.,  {Leh{\'a}r} J.,   {McLeod} B.~A.,  2006, \mn@doi [\nar]
  {10.1016/j.newar.2006.06.038}, \href
  {http://adsabs.harvard.edu/abs/2006NewAR..50..689P} {50, 689}

\bibitem[\protect\citeauthoryear{{Petrillo} et~al.,}{{Petrillo}
  et~al.}{2017}]{PetrilloEtal17}
{Petrillo} C.~E.,  et~al., 2017, \mn@doi [\mnras] {10.1093/mnras/stx2052},
  \href {http://adsabs.harvard.edu/abs/2017MNRAS.472.1129P} {472, 1129}

\bibitem[\protect\citeauthoryear{{Planck Collaboration} et~al.,}{{Planck
  Collaboration} et~al.}{2016}]{Planck16a}
{Planck Collaboration} et~al., 2016, \mn@doi [\aap]
  {10.1051/0004-6361/201525830}, \href
  {http://adsabs.harvard.edu/abs/2016A%26A...594A..13P} {594, A13}

\bibitem[\protect\citeauthoryear{Plous}{Plous}{1993}]{Plous93}
Plous S.,  1993, The Psychology of Judgment and Decision Making.
McGraw-Hill Education, \url {https://books.google.com/books?id=xvWOQgAACAAJ}

\bibitem[\protect\citeauthoryear{{Rathna Kumar} et~al.,}{{Rathna Kumar}
  et~al.}{2013}]{RathnaEtal13}
{Rathna Kumar} S.,  et~al., 2013, \mn@doi [\aap] {10.1051/0004-6361/201322116},
  \href {http://adsabs.harvard.edu/abs/2013A%26A...557A..44R} {557, A44}

\bibitem[\protect\citeauthoryear{{Refsdal}}{{Refsdal}}{1964}]{Refsdal64}
{Refsdal} S.,  1964, \mnras, \href
  {http://adsabs.harvard.edu/cgi-bin/nph-bib_query?bibcode=1964MNRAS.128..307R&db_key=AST}
  {128, 307}

\bibitem[\protect\citeauthoryear{{Riess} et~al.,}{{Riess}
  et~al.}{1998}]{RiessEtal98}
{Riess} A.~G.,  et~al., 1998, \mn@doi [\aj] {10.1086/300499}, \href
  {http://adsabs.harvard.edu/abs/1998AJ....116.1009R} {116, 1009}

\bibitem[\protect\citeauthoryear{{Riess} et~al.,}{{Riess}
  et~al.}{2016}]{RiessEtal16}
{Riess} A.~G.,  et~al., 2016, \mn@doi [\apj] {10.3847/0004-637X/826/1/56},
  \href {http://adsabs.harvard.edu/abs/2016ApJ...826...56R} {826, 56}

\bibitem[\protect\citeauthoryear{{Rusu} et~al.,}{{Rusu}
  et~al.}{2017}]{RusuEtal17}
{Rusu} C.~E.,  et~al., 2017, \mn@doi [\mnras] {10.1093/mnras/stx285}, \href
  {http://adsabs.harvard.edu/abs/2017MNRAS.467.4220R} {467, 4220}

\bibitem[\protect\citeauthoryear{{Schechter} \& {Wambsganss}}{{Schechter} \&
  {Wambsganss}}{2002}]{SchechterWambsganss02}
{Schechter} P.~L.,  {Wambsganss} J.,  2002, \mn@doi [\apj] {10.1086/343856},
  \href {http://adsabs.harvard.edu/abs/2002ApJ...580..685S} {580, 685}

\bibitem[\protect\citeauthoryear{{Schechter} et~al.,}{{Schechter}
  et~al.}{1997}]{SchechterEtal97}
{Schechter} P.~L.,  et~al., 1997, \mn@doi [\apjl] {10.1086/310478}, \href
  {http://adsabs.harvard.edu/abs/1997ApJ...475L..85S} {475, L85}

\bibitem[\protect\citeauthoryear{{Schechter}, {Morgan}, {Chehade}, {Metcalfe},
  {Shanks}  \& {McDonald}}{{Schechter} et~al.}{2017}]{SchechterEtal17}
{Schechter} P.~L.,  {Morgan} N.~D.,  {Chehade} B.,  {Metcalfe} N.,  {Shanks}
  T.,   {McDonald} M.,  2017, \mn@doi [\aj] {10.3847/1538-3881/aa6899}, \href
  {http://adsabs.harvard.edu/abs/2017AJ....153..219S} {153, 219}

\bibitem[\protect\citeauthoryear{{Schneider} \& {Sluse}}{{Schneider} \&
  {Sluse}}{2013}]{SchneiderSluse13}
{Schneider} P.,  {Sluse} D.,  2013, \mn@doi [\aap]
  {10.1051/0004-6361/201321882}, \href
  {http://adsabs.harvard.edu/abs/2013A%26A...559A..37S} {559, A37}

\bibitem[\protect\citeauthoryear{{Schneider} \& {Sluse}}{{Schneider} \&
  {Sluse}}{2014}]{SchneiderSluse14}
{Schneider} P.,  {Sluse} D.,  2014, \mn@doi [\aap]
  {10.1051/0004-6361/201322106}, \href
  {http://adsabs.harvard.edu/abs/2014A%26A...564A.103S} {564, A103}

\bibitem[\protect\citeauthoryear{{Shajib}, {Treu}  \& {Agnello}}{{Shajib}
  et~al.}{2018}]{ShajibEtal18}
{Shajib} A.~J.,  {Treu} T.,   {Agnello} A.,  2018, \mn@doi [\mnras]
  {10.1093/mnras/stx2302}, \href
  {http://adsabs.harvard.edu/abs/2018MNRAS.473..210S} {473, 210}

\bibitem[\protect\citeauthoryear{{Shakura} \& {Sunyaev}}{{Shakura} \&
  {Sunyaev}}{1973}]{ShakuraSunyaev73}
{Shakura} N.~I.,  {Sunyaev} R.~A.,  1973, \aap, \href
  {http://adsabs.harvard.edu/abs/1973A%26A....24..337S} {24, 337}

\bibitem[\protect\citeauthoryear{{Sluse} et~al.,}{{Sluse}
  et~al.}{2003}]{SluseEtal03}
{Sluse} D.,  et~al., 2003, \mn@doi [\aap] {10.1051/0004-6361:20030904}, \href
  {http://adsabs.harvard.edu/abs/2003A%26A...406L..43S} {406, L43}

\bibitem[\protect\citeauthoryear{{Suyu}}{{Suyu}}{2012}]{Suyu12}
{Suyu} S.~H.,  2012, ArXiv e-prints (1202.0287), \href
  {http://adsabs.harvard.edu/abs/2012arXiv1202.0287S} {}

\bibitem[\protect\citeauthoryear{{Suyu}, {Marshall}, {Auger}, {Hilbert},
  {Blandford}, {Koopmans}, {Fassnacht}  \& {Treu}}{{Suyu}
  et~al.}{2010}]{SuyuEtal10}
{Suyu} S.~H.,  {Marshall} P.~J.,  {Auger} M.~W.,  {Hilbert} S.,  {Blandford}
  R.~D.,  {Koopmans} L.~V.~E.,  {Fassnacht} C.~D.,   {Treu} T.,  2010, \mn@doi
  [\apj] {10.1088/0004-637X/711/1/201}, \href
  {http://adsabs.harvard.edu/abs/2010ApJ...711..201S} {711, 201}

\bibitem[\protect\citeauthoryear{{Suyu}, {Chang}, {Courbin}  \&
  {Okumura}}{{Suyu} et~al.}{2018}]{SuyuEtal18}
{Suyu} S.~H.,  {Chang} T.-C.,  {Courbin} F.,   {Okumura} T.,  2018, preprint,
  \href {http://adsabs.harvard.edu/abs/2018arXiv180107262S} {} (\mn@eprint
  {arXiv} {1801.07262})

\bibitem[\protect\citeauthoryear{{Tewes}, {Courbin}  \& {Meylan}}{{Tewes}
  et~al.}{2013a}]{TewesEtal13b}
{Tewes} M.,  {Courbin} F.,   {Meylan} G.,  2013a, \mn@doi [\aap]
  {10.1051/0004-6361/201220123}, \href
  {http://adsabs.harvard.edu/abs/2013A%26A...553A.120T} {553, A120}

\bibitem[\protect\citeauthoryear{{Tewes} et~al.,}{{Tewes}
  et~al.}{2013b}]{TewesEtal13a}
{Tewes} M.,  et~al., 2013b, \mn@doi [\aap] {10.1051/0004-6361/201220352}, \href
  {http://adsabs.harvard.edu/abs/2013A%26A...556A..22T} {556, A22}

\bibitem[\protect\citeauthoryear{{Tie} \& {Kochanek}}{{Tie} \&
  {Kochanek}}{2018}]{TieKochanek18}
{Tie} S.~S.,  {Kochanek} C.~S.,  2018, \mn@doi [\mnras]
  {10.1093/mnras/stx2348}, \href
  {http://adsabs.harvard.edu/abs/2018MNRAS.473...80T} {473, 80}

\bibitem[\protect\citeauthoryear{{Tihhonova} et~al.,}{{Tihhonova}
  et~al.}{2017}]{TihhonovaEtal17}
{Tihhonova} O.,  et~al., 2017, preprint, \href
  {http://adsabs.harvard.edu/abs/2017arXiv171108804T} {} (\mn@eprint {arXiv}
  {1711.08804})

\bibitem[\protect\citeauthoryear{{Tonry}}{{Tonry}}{1998}]{Tonry98}
{Tonry} J.~L.,  1998, \mn@doi [\aj] {10.1086/300170}, \href
  {http://adsabs.harvard.edu/abs/1998AJ....115....1T} {115, 1}

\bibitem[\protect\citeauthoryear{{Treu} \& {Koopmans}}{{Treu} \&
  {Koopmans}}{2002}]{TreuKoopmans02}
{Treu} T.,  {Koopmans} L.~V.~E.,  2002, \mn@doi [\mnras]
  {10.1046/j.1365-8711.2002.06107.x}, \href
  {http://adsabs.harvard.edu/abs/2002MNRAS.337L...6T} {337, L6}

\bibitem[\protect\citeauthoryear{{Treu} \& {Marshall}}{{Treu} \&
  {Marshall}}{2016}]{TreuMarshall17}
{Treu} T.,  {Marshall} P.~J.,  2016, \mn@doi [\aapr]
  {10.1007/s00159-016-0096-8}, \href
  {http://adsabs.harvard.edu/abs/2016A%26ARv..24...11T} {24, 11}

\bibitem[\protect\citeauthoryear{{Tsvetkova} et~al.,}{{Tsvetkova}
  et~al.}{2010}]{TsvetkovaEtal10}
{Tsvetkova} V.~S.,  et~al., 2010, \mn@doi [\mnras]
  {10.1111/j.1365-2966.2010.16882.x}, \href
  {http://adsabs.harvard.edu/abs/2010MNRAS.406.2764T} {406, 2764}

\bibitem[\protect\citeauthoryear{{Vegetti} \& {Koopmans}}{{Vegetti} \&
  {Koopmans}}{2009}]{VegettiKoopmans09}
{Vegetti} S.,  {Koopmans} L.~V.~E.,  2009, \mn@doi [\mnras]
  {10.1111/j.1365-2966.2008.14005.x}, \href
  {http://adsabs.harvard.edu/abs/2009MNRAS.392..945V} {392, 945}

\bibitem[\protect\citeauthoryear{{Wambsganss}}{{Wambsganss}}{2006}]{Wambsganss06}
{Wambsganss} J.,  2006, ArXiv Astrophysics e-prints, \href
  {http://adsabs.harvard.edu/abs/2006astro.ph..4278W} {}

\bibitem[\protect\citeauthoryear{{Warren} \& {Dye}}{{Warren} \&
  {Dye}}{2003}]{WarrenDye03}
{Warren} S.~J.,  {Dye} S.,  2003, \mn@doi [\apj] {10.1086/375132}, \href
  {http://adsabs.harvard.edu/cgi-bin/nph-bib_query?bibcode=2003ApJ...590..673W&db_key=AST}
  {590, 673}

\bibitem[\protect\citeauthoryear{{Weinberg}, {Mortonson}, {Eisenstein},
  {Hirata}, {Riess}  \& {Rozo}}{{Weinberg} et~al.}{2013}]{WeinbergEtal13}
{Weinberg} D.~H.,  {Mortonson} M.~J.,  {Eisenstein} D.~J.,  {Hirata} C.,
  {Riess} A.~G.,   {Rozo} E.,  2013, \mn@doi [\physrep]
  {10.1016/j.physrep.2013.05.001}, \href
  {http://adsabs.harvard.edu/abs/2013PhR...530...87W} {530, 87}

\bibitem[\protect\citeauthoryear{{Williams} et~al.,}{{Williams}
  et~al.}{2018}]{WilliamEtal18}
{Williams} P.~R.,  et~al., 2018, \mn@doi [\mnras] {10.1093/mnrasl/sly043},
  \href {http://adsabs.harvard.edu/abs/2018MNRAS.tmpL..44W} {}

\bibitem[\protect\citeauthoryear{{Wong} et~al.,}{{Wong}
  et~al.}{2017}]{WongEtal17}
{Wong} K.~C.,  et~al., 2017, \mn@doi [\mnras] {10.1093/mnras/stw3077}, \href
  {http://adsabs.harvard.edu/abs/2017MNRAS.465.4895W} {465, 4895}

\bibitem[\protect\citeauthoryear{{Xu}, {Sluse}, {Schneider}, {Springel},
  {Vogelsberger}, {Nelson}  \& {Hernquist}}{{Xu} et~al.}{2015}]{XuEtal15}
{Xu} D.,  {Sluse} D.,  {Schneider} P.,  {Springel} V.,  {Vogelsberger} M.,
  {Nelson} D.,   {Hernquist} L.,  2015, ArXiv:1507.07937, \href
  {http://adsabs.harvard.edu/abs/2015arXiv150707937X} {}

\bibitem[\protect\citeauthoryear{{de Grijs}, {Courbin},
  {Mart{\'{\i}}nez-V{\'a}zquez}, {Monelli}, {Oguri}  \& {Suyu}}{{de Grijs}
  et~al.}{2017}]{deGrijsEtal17}
{de Grijs} R.,  {Courbin} F.,  {Mart{\'{\i}}nez-V{\'a}zquez} C.~E.,  {Monelli}
  M.,  {Oguri} M.,   {Suyu} S.~H.,  2017, \mn@doi [\ssr]
  {10.1007/s11214-017-0395-z}, \href
  {http://adsabs.harvard.edu/abs/2017SSRv..212.1743D} {212, 1743}

\makeatother
\end{thebibliography}






\bsp	
\label{lastpage}
\end{document}